%% file: main.tex
\documentclass[10pt,letterpaper,twocolumn,conference]{IEEEtran}
\usepackage[utf8]{inputenc}
\usepackage[T1]{fontenc}
\usepackage{amsmath}
\usepackage{amsfonts}
\usepackage{amssymb}

\usepackage{graphicx}
\usepackage{xspace}
\usepackage[table]{xcolor}

\usepackage{hyperref}
\usepackage{caption}
\usepackage{subcaption}

\usepackage{booktabs}
\usepackage{tabularx}
\usepackage{multirow}
\usepackage{array}

\graphicspath{{./figures/}}

\renewcommand{\paragraph}[1]{\vspace{0.5em}\noindent\textbf{#1.}}

\usepackage{pifont}
\let\oldding\ding%
\renewcommand{\ding}[2][1]{\scalebox{#1}{\oldding{#2}}}
\newcommand{\one}{\ding[1.2]{192}\xspace}
\newcommand{\two}{\ding[1.2]{193}\xspace}
\newcommand{\three}{\ding[1.2]{194}\xspace}
\newcommand{\four}{\ding[1.2]{195}\xspace}
\newcommand{\five}{\ding[1.2]{196}\xspace}
\newcommand{\six}{\ding[1.2]{197}\xspace}
\newcommand{\seven}{\ding[1.2]{198}\xspace}
\newcommand{\eight}{\ding[1.2]{199}\xspace}

\newcolumntype{L}{>{\raggedright\let\newline\\\arraybackslash\hspace{0pt}}X}

\input{definitions.tex}

\title{Phishing in Organizations: Findings from a Large-Scale and Long-Term Study}

\author{\IEEEauthorblockN{Daniele Lain, Kari Kostiainen, and Srdjan \v{C}apkun}
	\IEEEauthorblockA{\itshape Department of Computer Science\\
		ETH Zurich, Switzerland\\
	\normalfont \{daniele.lain, kari.kostiainen, srdjan.capkun\} @inf.ethz.ch}
}

\begin{document}
\maketitle

\begin{abstract}
	\input{sections/abstract.tex}
\end{abstract}

\section{Introduction}
\label{sec:introduction}
\input{sections/introduction.tex}

\section{Research Questions and Main Findings}
\label{sec:findings}
\input{sections/findings.tex}

\section{Experimental Setup}
\label{sec:exp_setup}
\input{sections/setup.tex}

\section{Which Employees Fall for Phishing?}
\label{sec:results_demo}
\input{sections/results_demo.tex}

\section{Phishing Vulnerability Over Time}
\label{sec:results_time}
\input{sections/results_time.tex}

\section{Effectiveness of Warnings and Training}
\label{sec:results_tools}
\input{sections/results_tools.tex}

\section{Can Employees Help the Organization?}
\label{sec:results_ucp2}
\input{sections/results_ucp2.tex}

\section{Study Validity}
\label{sec:discussion}
\input{sections/discussion_limitations.tex}

\section{Related Work}
\label{sec:relwork}

\input{sections/related_work.tex}

\section{Conclusions and Future Work}
\label{sec:conclusion}
\input{sections/conclusion.tex}

\section*{Acknowledgments}
This research has been partially supported by the Zurich Information Security and Privacy Center (ZISC).

\bibliographystyle{IEEEtran}
\bibliography{bibliography}

\appendix
\section{Appendix}
\label{app:additional_results}
\input{sections/additional_results.tex}

\end{document}

%% file: definitions.tex
\newcommand{\numusers}{14,733\xspace}
\newcommand{\numbuttons}{21,000\xspace}
\newcommand{\numemployees}{56,000\xspace}

\newcommand{\minage}{18\xspace}
\newcommand{\maxage}{73\xspace}

\newcommand{\smallerGroup}{1,223\xspace}
\newcommand{\largerGroup}{1,231\xspace}
\newcommand{\numEmails}{8\xspace}

\newcommand{\numquestionnaire}{151\xspace}

\newcommand{\RQ}{\texttt{RQ1}\xspace}
\newcommand{\RQQ}{\texttt{RQ2}\xspace}
\newcommand{\RQQQ}{\texttt{RQ3}\xspace}
\newcommand{\RQQQQ}{\texttt{RQ4}\xspace}

%% file: sections/abstract.tex
In this paper, we present findings from a large-scale and long-term phishing experiment that we conducted in collaboration with a partner company. Our experiment ran for 15 months during which time more than 14,000 study participants (employees of the company) received different simulated phishing emails in their normal working context. We also deployed a reporting button to the company's email client which allowed the participants to report suspicious emails they received. We measured click rates for phishing emails, dangerous actions such as submitting credentials, and reported suspicious emails.

The results of our experiment provide three types of contributions. First, some of our findings support previous literature with improved ecological validity. One example of such results is good effectiveness of warnings on emails. Second, some of our results contradict prior literature and common industry practices. Surprisingly, we find that embedded training during simulated phishing exercises, as commonly deployed in the industry today, does not make employees more resilient to phishing, but instead it can have unexpected side effects that can make employees even more susceptible to phishing. 
And third, we report new findings. In particular, we are the first to demonstrate that using the employees as a collective phishing detection mechanism is practical in large organizations. Our results show that such crowd-sourcing allows fast detection of new phishing campaigns, the operational load for the organization is acceptable, and the employees remain active over long periods of time.

%% file: sections/introduction.tex
Phishing remains a major problems on the Internet~\cite{verizon2020dbir}.
Deceptive emails that trick users to perform unsafe actions are getting increasingly sophisticated~\cite{verizon2019dbir,verizon2020dbir} and during the last two decades phishing showed no sign of slowing down~\cite{hong2012state}. 
The job of cyber-criminals is made easy by the development of \emph{phishing kits}, software capable of automatically creating deceptive copies of popular websites~\cite{oest2018inside,cova2008there,han2016phisheye}.
To make things even worse, the COVID-19 pandemic has shifted work, shopping and other activities online which in turn has created new phishing opportunities and increased phishing~\cite{bitaab2020scam}.

Researchers have studied phishing for decades (see~\cite{khonji2013phishing,aleroud2017phishing,almomani2013survey,jampen2020don} for extensive reviews of early works) and proposed various defenses from email filters ~\cite{almomani2013survey, fette2007learning}, to detection of phishing websites~\cite{xiang2011cantina+}, patterns of phishing campaigns~\cite{oest2020sunrise}, triggers that push people to fall for phishing~\cite{wang2012research}, and ways to educate users~\cite{wash2018provides,jampen2020don}.
During the last decade, also an entire ecosystem of companies that provide phishing prevention products and services has emerged. Common commercial offerings include training and educational services~\cite{wombattraining,cofense,rapid7,knowbe4}, databases of known URLs as well as emails used by phishing attacks~\cite{phishtank,openphish,moore2008evaluating}, and email filters powered by threat intelligence collected by specialists and reports from customers~\cite{wombatbutton,cofense,rapid7}.

\paragraph{Our study and contributions}
In this paper, we study phishing with a particular focus on phishing in \emph{organizations}. 
We approach this topic through the following four questions -- all related to human factors of phishing.
First, we are interested to understand \emph{which employees are the most vulnerable} to phishing in large organizations. 
We examine this through common aspects like employee demographics and job type.
Second, we explore how the organization's \emph{phishing vulnerability evolves over time}.
For instance, we study how many employees will eventually fall for phishing in continued exposure to phishing.
Third, we study \emph{how organizations can help their employees} in phishing prevention. In particular, we analyze the benefits of currently popular tools such as embedded phishing training and warnings on top of suspicious emails.
And fourth, we explore whether \emph{the employees can collectively help the organization} in phishing prevention. 
Regarding this question, we focus on using the employees as a collective phishing detection sensor -- an idea that has been previously suggested~\cite{burda2020don, moore2008evaluating}, but prior to our work, its effectiveness and feasibility has not been publicly evaluated in a real large organization.

To answer these questions, we designed and conducted a large-scale and long-term phishing study in collaboration with a partner company. Our study ran for 15 months (July 2019-October 2020) and during it 14,773 
employees of the company became participants in our experiment. 
Our study involved sending simulated phishing emails to the participants, who received them as part of their normal work flow and context. We measured their click rates, submission of credentials, and enabling macros on attachments. 
We also deployed a reporting button to the corporate email client which allowed our study participants to easily report emails that they found suspicious, and analyzed the reported emails. 

To the best of our knowledge, our experiment is the first study of phishing in organizations that is at the same time large-scale (14k participants), long-term (15 months), realistic (we measure real employees' phishing behavior in their actual working context), and diverse (including participants across various corporate departments and job roles). All comparable, previous studies are either smaller~\cite{burda2020testing,burns2019spear,carella2017impact,gavett2017phishing,iuga2016baiting,kumaraguru2007getting,kumaraguru2007protecting,kumaraguru2008lessons,kumaraguru2009school,lin2019susceptibility,oliveira2017dissecting,petelka2019put,volkamer2017user,wash2018provides,williams2018exploring}, shorter~\cite{oliveira2017dissecting,lin2019susceptibility,burda2020testing,wash2018provides,kumaraguru2008lessons,burns2019spear}, based on role-play~\cite{sheng2010falls,blythe2011f,gavett2017phishing,kumaraguru2007protecting,kumaraguru2007getting}, or less diverse~\cite{blythe2011f,kumaraguru2009school,gavett2017phishing,kumaraguru2007getting,carella2017impact,burns2019spear}, as we will elaborate in Section~\ref{sec:findings}.

The results of our experiment provide three types of contributions. 
First, we report several results that \emph{support} previous literature with increased ecological validity (e.g., more study participants, longer study duration, or more realistic study setting). 
Among others, we find email warnings are effective and observe many ``repeated clickers''~\cite{canham2019enduring} in our experiment.

Second, our study uncovers a few findings that \emph{contradict} both the conclusions of previous academic studies and common industry practices. In particular, we find that embedded phishing training, as commonly used in the industry today, can lead to unexpected side effects and even be detrimental to phishing prevention. This is a significant finding, due to wide use of this practice in the industry.

And third, our results provide \emph{new insights} to phishing in organizations. In particular, as one of the main contributions of this paper, our experiment is the first to demonstrate that crowd-sourced phishing detection can be effective, fast, and sustainable over long periods of time. During our experiment, the employees reported thousands of suspicious emails which represented hundreds of real and previously unseen phishing campaigns. The reporting speed of our simulated phishing emails indicates that new campaigns can be detected within few minutes from their launch. We designed a simple processing pipeline that combined automated and manual analysis for the reported emails. Our experiment shows that through such techniques, the operational load of handling all the reported emails can be made low even in large organizations. Our experiment also demonstrates that large employee bases can collectively retain sufficiently high reporting rates over long periods of time. In summary, this paper is the first to demonstrate that crowd-sourced phishing detection is a practical and effective option for many organizations.

To summarize, this paper makes the following contributions:

\begin{enumerate}
    \item \emph{Extensive measurement study} on human factors of phishing and phishing prevention in large organizations. 
    \item \emph{Supportive results} for several previous research findings with improved ecological validity. 
    \item \emph{Contradicting findings} that challenge the conclusions of previous research studies and popular industry practices.
    \item \emph{Large-scale evaluation of crowd-sourced phishing reporting} that shows fast detection, small operational overhead, and sustained employee reporting activity.
\end{enumerate}

\paragraph{Paper outline}
This paper is organized as follows. In Section~\ref{sec:findings}, we define our research questions and provide an overview of our findings.
We describe our experimental setup in Section~\ref{sec:exp_setup}.
We report results related to employee demographics in Section~\ref{sec:results_demo}. Section~\ref{sec:results_time} shows how phishing vulnerability evolved over time in our study.
Section~\ref{sec:results_tools} explains our results related to warnings and embedded training. Section~\ref{sec:results_ucp2} analyzes crowd-sourced phishing detection.
In Section~\ref{sec:discussion} we discuss validity of our study. Section~\ref{sec:relwork} reviews related work and Section~\ref{sec:conclusion} concludes the paper.

%% file: sections/findings.tex
\begin{table*}[th]
	\renewcommand{\arraystretch}{2}
	\centering
	\caption{\textbf{Summary of our research questions and main results} that include findings that support prior literature, findings that contradict previous studies, and new insights. The two most significant contributions of this paper are \textbf{marked in bold}.}
	\rowcolors{2}{white}{gray!10}
	\begin{tabularx}{\linewidth}{@{}LXLX@{}}
		\toprule  
		& \textbf{Findings that support} previous studies in literature
		& \textbf{Findings that contradict} previous studies in literature
		& \textbf{New findings} on phishing in organizations
		\\[1.5em] 
		\midrule
		
		\textbf{\RQ: Which employees fall for phishing?} (Section~\ref{sec:results_demo})
		& Age and computer skills correlate with phishing susceptibility~\cite{sheng2010falls,blythe2011f,kumaraguru2009school,oliveira2017dissecting,lin2019susceptibility,gavett2017phishing,jampen2020don,burda2020testing} 
		& Gender does not correlate with phishing susceptibility {\scriptsize(contradicts~\cite{oliveira2017dissecting,iuga2016baiting,sheng2010falls,blythe2011f})}
		&  Type of computer use is more predictive for phishing vulnerability than amount of computer use \\
		
		\textbf{\RQQ: How does organization's vulnerability to phishing evolve over time?} (Section~\ref{sec:results_time})
		& There are several ``repeated clickers'' in a large organization~\cite{canham2019enduring} 
		&  
		& Many employees will eventually fall for phishing if continuously exposed \\
		
		\textbf{\RQQQ: How effective are phishing warnings and training?} (Section~\ref{sec:results_tools})
		& Warnings on top of suspicious emails are effective~\cite{volkamer2017user,petelka2019put,williams2018exploring}
		& \textbf{Voluntary embedded training in simulated phishing exercises is not effective} {\scriptsize(contradicts~\cite{kumaraguru2008lessons, kumaraguru2007protecting,kumaraguru2009school})}
		& More detailed warnings are not more effective than simple ones
		\\[2.5em]
		
		\textbf{\RQQQQ: Can employees help the organization in phishing detection?} (Section~\ref{sec:results_ucp2})
		&  
		&  
		&  \textbf{Crowd-sourcing phishing email detection is both effective and feasible}
		\\[1.5em]
		\bottomrule
	\end{tabularx}
	\label{tab:findings}
\end{table*}

In this section, we first define the research questions that our study was designed to answer and then provide a summary of our main findings, both summarized in Table~\ref{tab:findings}.

\subsection{Research Questions}
\label{sec:rqs}


\vspace{.7em}\noindent\textbf{\RQ: Which employees fall for phishing?} 
The first goal of our experiment was to understand which employees in a large organization are the most likely to fall for phishing. 
In particular, we wanted to understand how employee characteristics that are easily available to organizations, such as age, gender, and the assumed level of computer use in one's job type, correlate with phishing susceptibility. 

\vspace{.7em}\noindent\textbf{\RQQ: How does organization's vulnerability to phishing evolve over time?} 
The second goal of our experiment was to understand how the continued presence of phishing affects organizations over time. We examine topics like how large is the fraction of the employee base that will eventually fall for phishing, and how many individuals repeatedly fall for phishing~\cite{canham2019enduring}. 

\vspace{.7em}\noindent\textbf{\RQQQ: How effective are phishing warnings and training?}
Our third goal was to understand how large organizations can help their employees to recognize phishing emails and thus defend themselves against phishing. 
Today, organizations can choose from a range of tools and educational measures designed for this purpose. 
In our study, we focused on evaluating tools that can be deployed to a large employee base with a moderate cost, as such tools are commonly used in practice.

The first tool whose effectiveness we decided to examine was \emph{warnings} on top of suspicious emails. 
Warnings are used in many popular email clients and services such as Gmail~\cite{gmailwarnings}: they are shown on top of the emails where automated phishing detection mechanism has identified some risky or suspicious features in the email, but it could not label the email as phishing with sufficiently high confidence (often email filters are tuned to be permissive to avoid too many false positives). 

The second tool we wanted to test was \emph{simulated phishing exercises}~\cite{kumaraguru2007protecting,jampen2020don} in combination with \emph{embedded training}~\cite{kumaraguru2008lessons}.
During the last decade, simulated phishing exercises have become a common industry practice~\cite{cofense,rapid7,wombattraining,knowbe4}. 
In a simulated phishing exercise, the organization sends emails that mimic real phishing emails to their employees and then track which employees perform unsafe actions such as clicking links or disclosing credentials to a web page. 
Often such exercises are combined with embedded training (sometimes also called contextual training), where employees that fail the exercise (e.g., by clicking on a link or disclosing their credentials) are forwarded to an information resource like a web page that provides educational material about phishing.

\vspace{.7em}\noindent\textbf{\RQQQQ: Can employees help the organization in phishing detection?}
The fourth goal of our experiment was to understand if a large employee base can collectively help the organization in phishing prevention. 
More precisely, we wanted to understand whether using employees as a \emph{crowd-sourced phishing detection} mechanism is efficient (can phishing campaigns be detected fast enough?), practical (does the administrative load of processing the reported emails remain acceptable?), and sustainable (will employees continue to report emails over time?) in a large organization.
Additionally, our aim was to understand if the presence of a feedback mechanism encourages employees to report suspicious emails more. 


\subsection{Summary of Main Findings}
\label{sec:summary_findings}

Next, we provide an overview of the main findings of our experiment, and briefly discuss how these results relate to prior research literature (a more detailed survey of related work is given in Section~\ref{sec:relwork}). 
Our findings \emph{support} claims of previous studies with improved ecological validity; \emph{contradict} prior conclusions and common industry practices; and provide \emph{new insights} related to phishing in large organizations.


\paragraph{Findings related to \RQ}
The results of our experiment support previous work which has showed that age~\cite{sheng2010falls,blythe2011f,kumaraguru2009school,oliveira2017dissecting,lin2019susceptibility,gavett2017phishing} and computer skills~\cite{jampen2020don,burda2020testing} both correlate with phishing vulnerability. Similar to previous studies, we also find that older and younger employees are more at risk, as well as people with lower computer skills.
Our experiment improves the ecological validity of these studies which were either smaller~\cite{oliveira2017dissecting,lin2019susceptibility,iuga2016baiting,kumaraguru2009school,gavett2017phishing,burda2020testing}, shorter in duration~\cite{oliveira2017dissecting,lin2019susceptibility,burda2020testing}, featured populations with less diversity (e.g., mostly university students and employees~\cite{blythe2011f,kumaraguru2009school,gavett2017phishing}, skewed in age~\cite{lin2019susceptibility}), or featured role-play or quiz-style studies only~\cite{sheng2010falls,blythe2011f,gavett2017phishing}.
Opposed to previous literature~\cite{oliveira2017dissecting,iuga2016baiting,sheng2010falls,blythe2011f}, we do not find gender to correlate with phishing susceptibility. The correlation that we observe is explained much better by skewed distribution of the different types of jobs among genders. We improve on these studies by reporting on a larger and more diverse population in their day-to-day job environment.

As a new finding, our study shows that the most vulnerable employees are those who use computers daily for repetitive task with a specialized software only, rather than those employees who do not need computers in their day-to-day job. That is, in our experiment, the \emph{type} of computer use is more predictive for phishing vulnerability than the \emph{amount}. We discuss these topics more in Section~\ref{sec:results_demo}.

\paragraph{Findings related to \RQQ}
Similar to previous studies, we find several ``repeated clickers'' who fail simulated phishing exercises multiple times~\cite{canham2019enduring}. 
We also find that if exposure to phishing continues in an organization, eventually a significant fraction of employees will fall for phishing. 
We elaborate on these results in Section~\ref{sec:results_time}.

\begin{figure*}[ht]
	\centering
	\includegraphics[width=.7\linewidth]{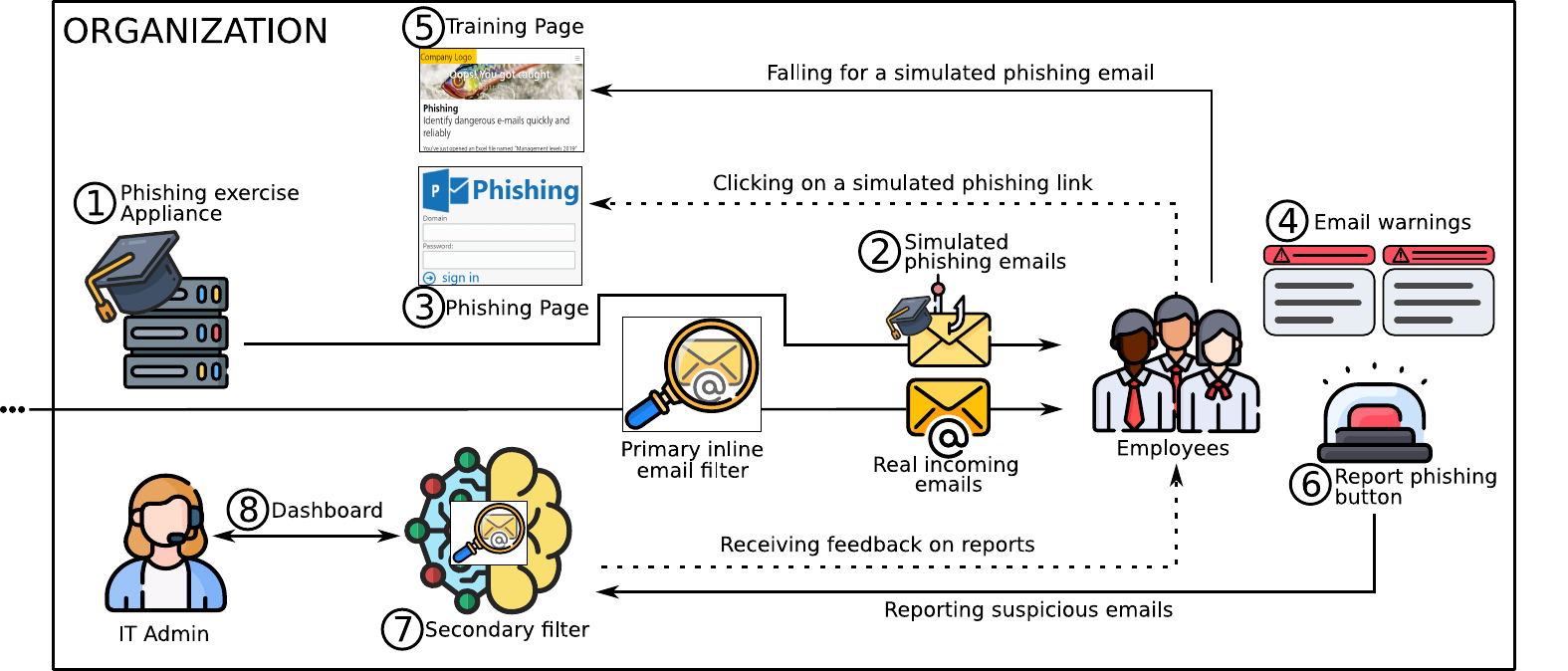}
	\caption{Overview of the measurement infrastructure that we deployed in the partner company.}
	\label{fig:infrastructure}
\end{figure*}

\paragraph{Findings related to \RQQQ}
Our results support the previous studies that find contextual warnings effective~\cite{volkamer2017user,petelka2019put,williams2018exploring} and the common industry practice of using such warnings~\cite{gmailwarnings}.
We improve these studies thanks to a larger, more general population with a rigorous control group.

Interestingly, contradicting prior research results~\cite{kumaraguru2008lessons, kumaraguru2007protecting,kumaraguru2009school} and a common industry practice~\cite{rapid7,wombattraining,knowbe4,cofense}, we found that the combination of simulated phishing exercises and voluntary embedded training (i.e., employees were not required to complete the training) not only failed to improve employee's phishing resilience, but it actually even the made employees more susceptible to phishing. 
Compared to our experiment, previous studies featured less participants~\cite{kumaraguru2008lessons,kumaraguru2007getting,carella2017impact,wash2018provides,kumaraguru2007protecting,burns2019spear}, were shorter in time~\cite{wash2018provides,kumaraguru2008lessons,burns2019spear}, had populations with little diversity~\cite{kumaraguru2007getting,carella2017impact,burns2019spear} or tested a role-playing setting only~\cite{kumaraguru2007protecting,kumaraguru2007getting}.
Our results suggest caution in the design of embedded training: we discuss the possible reasons (such as false sense of corporate IT security) and practical implications of this somewhat surprising and non-intuitive finding at length in Section~\ref{sec:results_tools}.

Another novel finding of our study is that adding more details to contextual warnings (e.g., explaining the reasons the email was flagged as suspicious) does not reduce phishing effectiveness significantly.

\paragraph{Findings related to \RQQQQ}
One of the main contributions of this paper is that we demonstrate experimentally that crowd-sourced phishing detection can be efficient and sustainable in large organizations. The idea of crowd-sourcing phishing detection to employees has been suggested in previous papers~\cite{burda2020don, moore2008evaluating}. Our contribution is that we are the first to evaluate this idea over a long period of time in the context of a real large organization.\footnote{Phishing reporting by users is a widely-used industry practice. For example, there are service providers who aggregate data from many of their business customers~\cite{wombatbutton,rapid7} and large email providers who collect reports to feed machine learning models~\cite{gmailreport}. However, prior to our work, it has not been publicly evaluated whether the employee base of a \textit{single} organization can be effectively leveraged as a phishing detection mechanism.} 
Our experiment shows that crowd-sourced phishing detection enables organizations to detect a large number of previously unseen real phishing campaigns with a short delay from the start of the campaign. The processing pipeline that we developed as part of our experiment also shows that the operational load of phishing report processing can be kept small, even in large organizations. Our study also demonstrates that a sufficiently high number of employees report suspicious emails actively over long periods of time. In summary, we show that crowd-sourced phishing detection provides a viable option for many organizations.  Section~\ref{sec:results_ucp2} provides full discussion of this topic.

%% file: sections/setup.tex
In this section we explain how we performed this study in collaboration with a partner company.

\subsection{Study Organization}
\label{sec:study_organization}

\paragraph{Partner company}
For this study, we collaborated with a company which employs more than \numemployees people of diverse technical skills, age groups, and jobs.
Our partner company is a large public company, dealing in logistics, finance, transport, and IT services. They employ people with different duties: field workers, branch workers that work in front-end stores in contact with the general public, and office workers of different qualifications, from IT to marketing and accounting.

As is common industry practice~\cite{rapid7,wombattraining,knowbe4,cofense}, at the time of study planning our partner company was already running a phishing awareness campaign 
which included simulated phishing emails and contextual (embedded) training.

\paragraph{Our role}
For this study, we leveraged the already existing phishing awareness campaign as a testbed for our research questions.
More precisely, we collaborated with our partner company in two ways.
First, as a scientific advisor who helped in the design of the experiment. By deploying different tools and conditions to different employees, we were able to use the existing campaign to address our research questions.
At the end of the study, we received anonymized data from the company in bulk and analyzed it.
Second, we took an active part by administering a questionnaire to randomly-selected employees and analyzing the responses.

\paragraph{Company's role}
The role of the company was three-fold: it designed all the simulated phishing emails; provided the infrastructure for sending the simulated phishing emails and measuring dangerous actions such as clicks; and hosted the embedded training resources (an educational webpage that was shown to those employees who performed a dangerous action).
The company had a pre-existing collaboration with an external service provider that specializes in phishing awareness and education. This service provider assisted the company in phishing email and contextual training page design.
The study was initiated and approved by the CISO of the company.

\subsection{Measurement Infrastructure}
\label{sec:infrastructure}

\paragraph{Phishing exercise component}
Our partner company deployed a phishing exercise component, shown as~\one in Figure~\ref{fig:infrastructure}, implemented by the service provider who specializes in phishing awareness and training, that sent simulated phishing emails~\two crafted by human experts.
These emails could either link to a deceptive website (hosted by component~\three) or have a malicious file attached, with the goal of deceiving the participant to do a \textit{dangerous action}, such as submitting their credentials or enabling macros on an attachment.

\begin{figure}[t]
	\centering
	\begin{subfigure}[b]{0.85\linewidth}
		\centering
		\includegraphics[width=\linewidth]{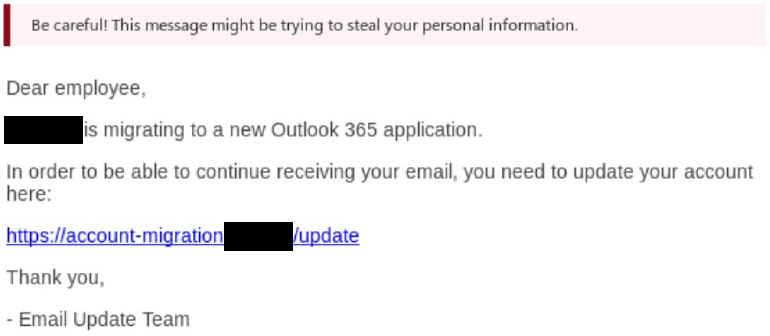}
		\caption{Short warnings.}
		\label{fig:warnings_short}
	\end{subfigure}
	~
	\begin{subfigure}[b]{0.85\linewidth}
		\centering
		\includegraphics[width=\linewidth]{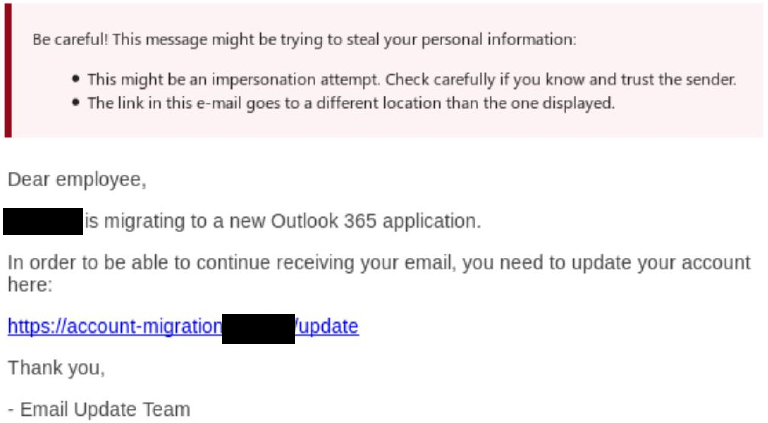}
		\caption{Detailed warnings.}
		\label{fig:warnings_long}
	\end{subfigure}
	\caption{Warnings that we added to selected participants' email clients on top of simulated phishing emails.}
	\label{fig:warnings}
\end{figure}

\paragraph{Deployed warnings}
Based on our recommendation, the  company deployed two types of warnings~\four that could be triggered to appear on top of the simulated phishing emails on the employees' email client (Outlook).
As a baseline, we deployed \textit{short} warnings (Figure~\ref{fig:warnings_short}), visually identical to the standard Outlook warnings that employees are used to, containing a similar generic sentence, warning the recipient to be careful because the email ``looks suspicious''.
	
We also developed and deployed \textit{detailed} warnings, shown in Figure~\ref{fig:warnings_long}, again visually identical to Outlook warnings, but adding a list of reasons why the email might be suspicious, e.g., mismatches between the email of the sender and the displayed name, or mismatches between the displayed link and the pointed domain.
Such information could be generated automatically in a deployment that adds warnings to emails that seem suspicious, when there is not sufficient certainty to block the email.

\begin{figure}[t]
	\centering
	\includegraphics[width=0.95\linewidth]{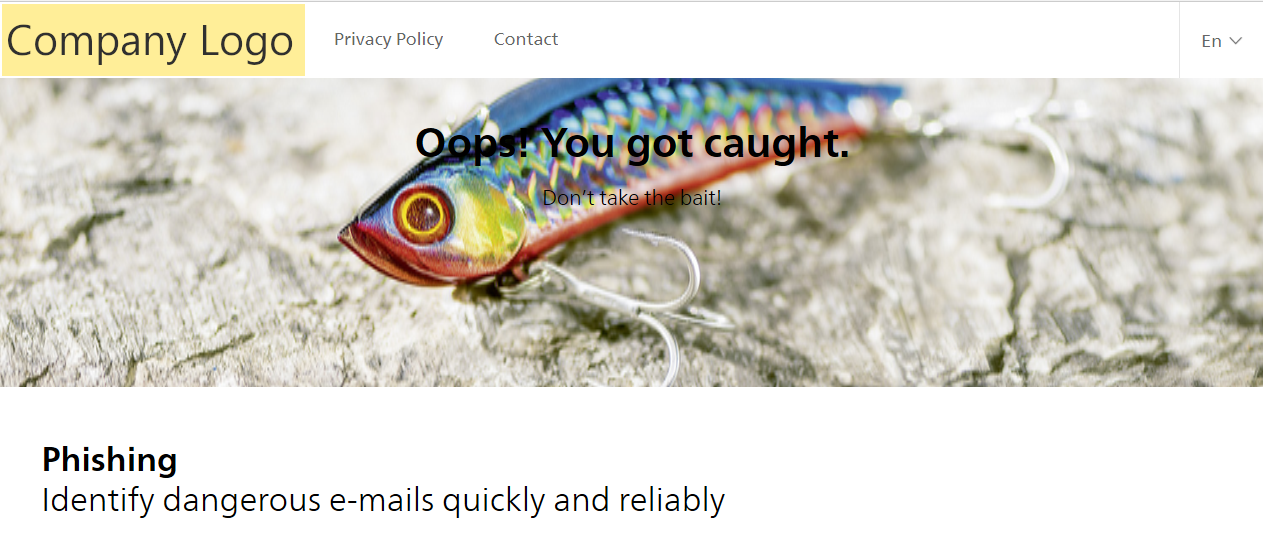}
	\caption{Header of the contextual training awareness web page that was displayed after falling for a simulated phishing email.}
	\label{fig:awareness}
\end{figure}

\paragraph{Deployed training}
The phishing exercise component also hosted a training web page on phishing~\five 
shown after someone performed the dangerous action of a simulated phishing email.
This internal corporate web page (part of it shown in  Figure~\ref{fig:awareness}) explained to the employee what happened in detail (i.e., that they failed a phishing exercise from their organization), specific cues one should have paid attention to in the email, tips to avoid phishing in the future, an instructional video, and further quizzes and learning material on phishing.
The training page was developed according to the best practices in academia~\cite{kumaraguru2010teaching,kumaraguru2007getting} and industry~\cite{rapid7,wombattraining} by the external service provider; we provide an excerpt in Appendix~\ref{app:training}.
The training page was delivered to employees such that there was no enforcement that the employee has to read the whole webpage or take the quizzes.

\begin{figure}[t]
	\centering
	\includegraphics[width=0.95\linewidth]{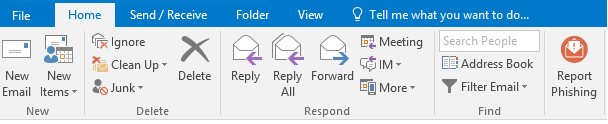}
	\caption{Menu bar of the company's email client (Outlook), modified to include a button to report suspicious emails.}
	\label{fig:button}
\end{figure}

\paragraph{Reporting button}
Our partner company deployed a button~\six for reporting suspicious emails. 
This button was introduced in the Outlook client, as shown in Figure~\ref{fig:button}, and it was advertised in the internal news of the company before the start of the experiment.
When reporting a suspicious email, employees could toggle a checkbox to report that they also opened the attachment or visited the link in the email, to notify the IT department about a possible incident.

\paragraph{Reported email processing}
All emails that were reported by our study participants were triaged by a commercial anti-phishing appliance~\seven that ran a more-detailed secondary analysis. The secondary analysis performed on the reported emails differed from the company's primary inline filter in two  ways: (a) more time consuming checks, such as following links, were performed, and (b) the analysis settings were tuned to be more aggressive, as at this point we did not need to avoid too many false positives. 
The results of the secondary analysis were presented to the company's IT department via a dashboard, where the appliance verdicts could be either confirmed or subverted based on manual analysis~\eight.
Further, the appliance could return feedback to the employees, indicating whether the reported email was indeed malicious or not. 

\subsection{Study Participants}
\label{sec:simulated_phishing}

\begin{figure*}[t]
	\centering
	\begin{subfigure}[b]{0.32\linewidth}
		\centering
		\includegraphics[width=\linewidth]{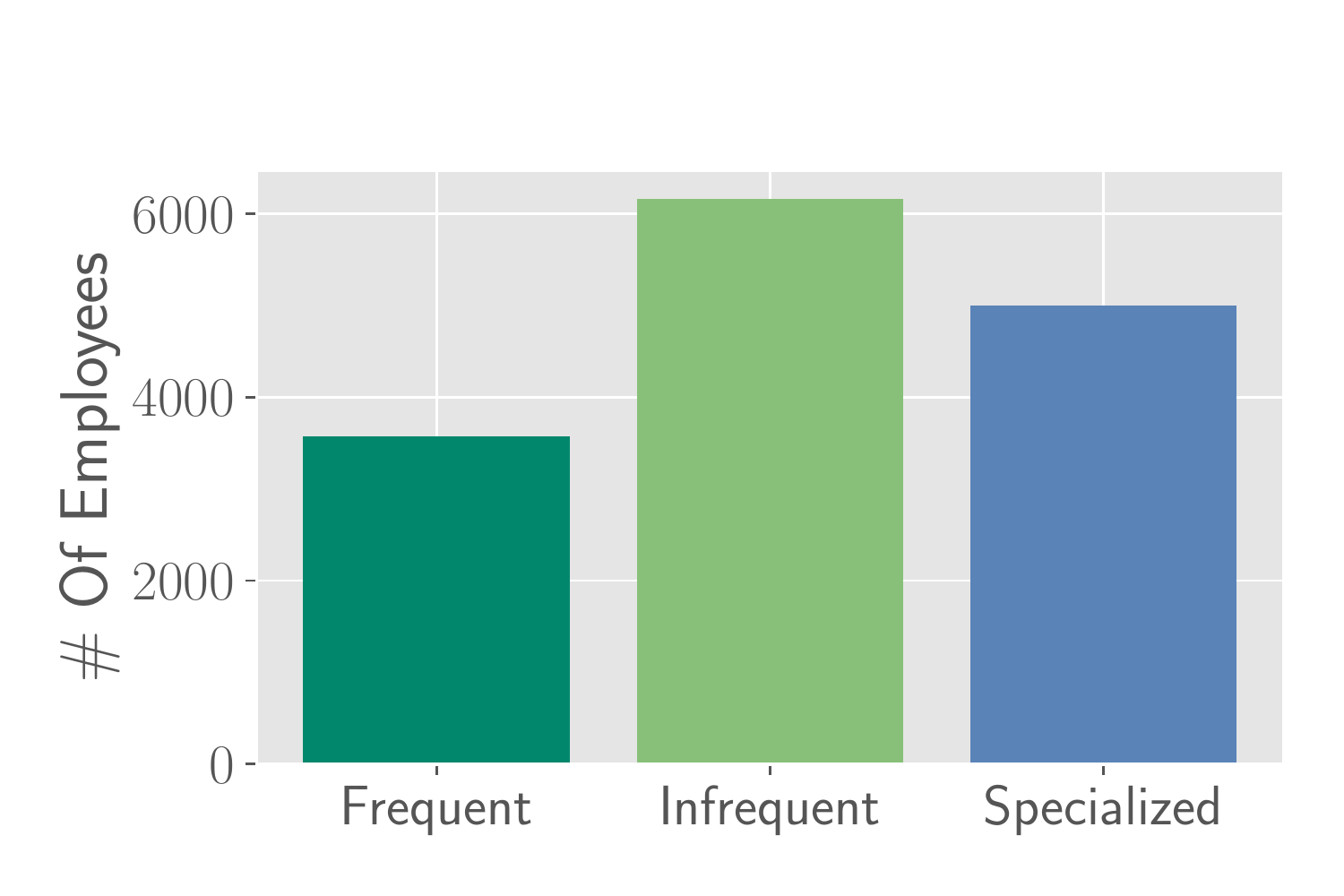}
		\caption{Use of computers in job.}
		\label{fig:setup_users_tech}
	\end{subfigure}
	\hfill
	\begin{subfigure}[b]{0.32\linewidth}
		\centering
		\includegraphics[width=\linewidth]{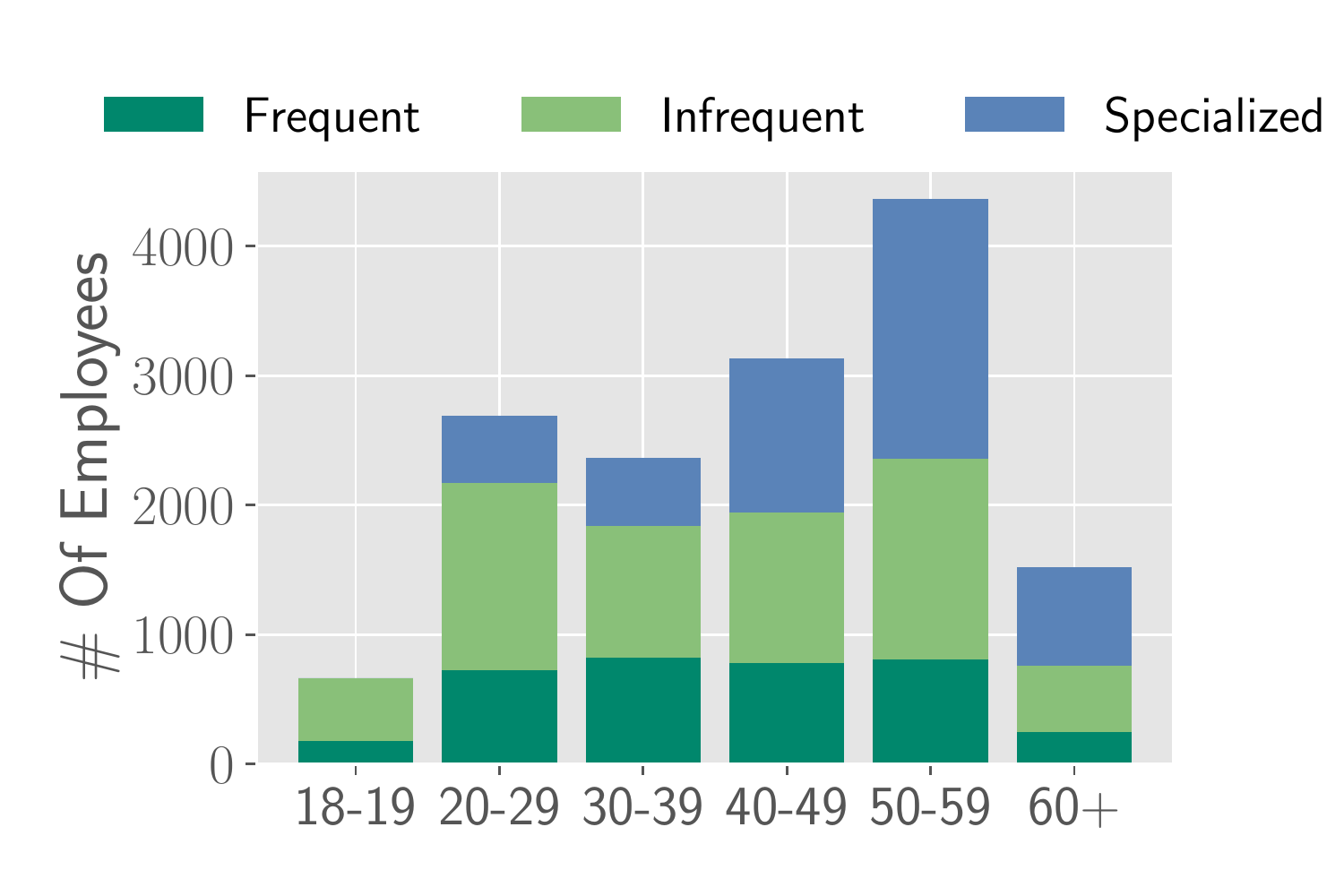}
		\caption{Age range.}
		\label{fig:setup_users_age}
	\end{subfigure}
	\hfill
	\begin{subfigure}[b]{0.32\linewidth}
		\centering
		\includegraphics[width=\linewidth]{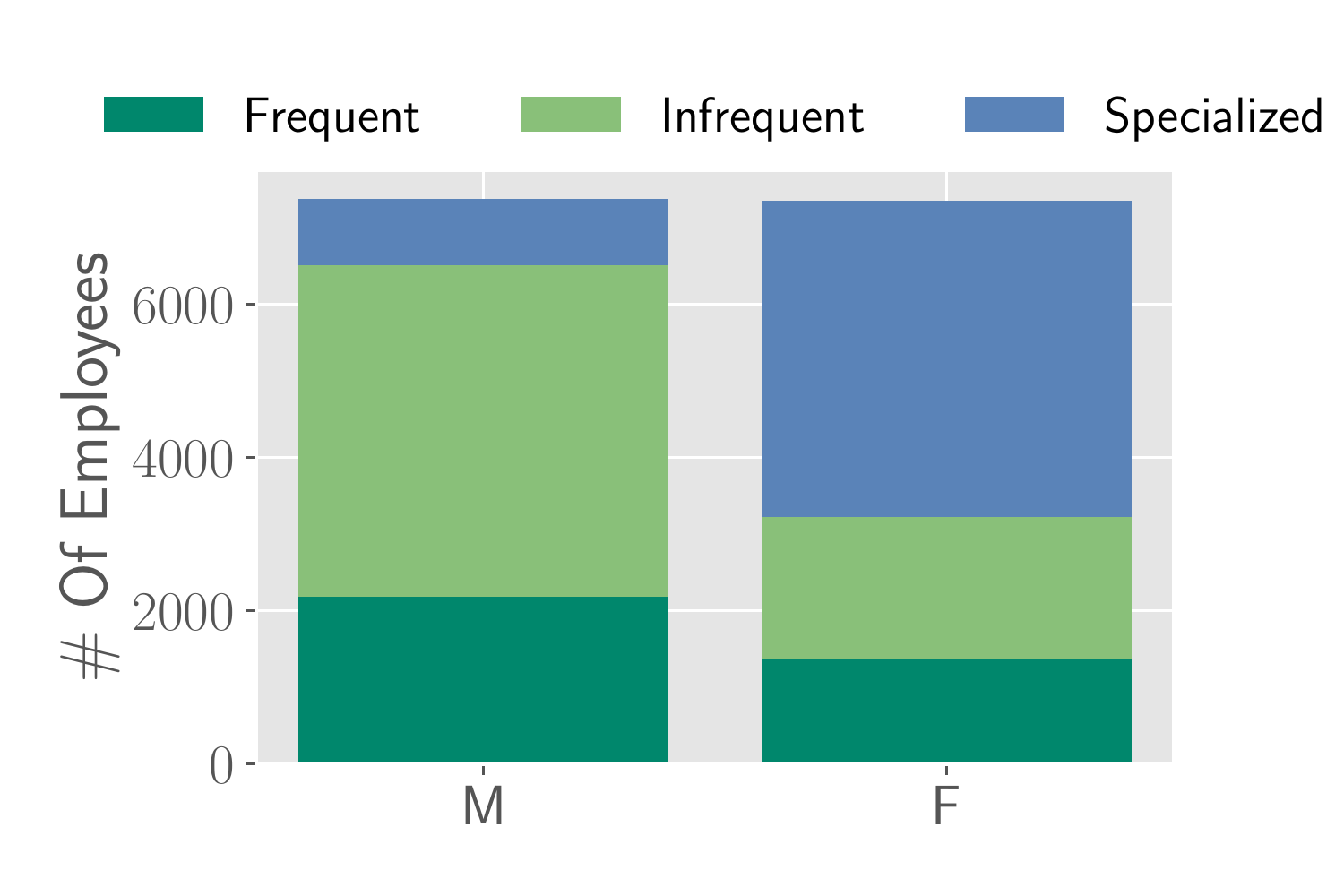}
		\caption{Gender.}
		\label{fig:setup_users_gender}
	\end{subfigure}
	\caption{Demographics information of the study participants. Age and gender are further divided by computer use in their job.}
	\label{fig:setup_users}
\end{figure*}

The company enrolled \numusers employees to be part of the experiment: we refer to them as \textit{participants}.
Participants were selected uniformly at random from the whole company's employee base, comprising many different job types, from accountants, IT, marketing, and managerial roles, to less technical jobs (e.g., in logistics, or working in retail shops). 
In total, participants spanned 28 organizational groups of the company, and represented 3,827 different teams.

For the purposes of our study, we classified the participants in terms of their age, gender and computer use in the day-to-day job. 
Participants were divided by computer use (shown in Figure~\ref{fig:setup_users_tech}) in three different categories: (i) office workers using computers daily, either in IT-related jobs or in jobs that use computers such as marketing and accounting (\textit{Frequent use}); (ii) employment roles such as retail shop workers in contact with the general public, that mostly use a point-of-sales software and configure services from it (\textit{Specialized use}); and (iii) roles such as team leaders of field workers in logistics who have a corporate email account but rarely use computers in their duties (\textit{Infrequent use}).
Study participants spanned an age range \minage-\maxage, with all age groups being well represented in our set of participants (Figure~\ref{fig:setup_users_age}). The gender distribution (Figure~\ref{fig:setup_users_gender}) of our participants was balanced: 7,377 were male, and 7,356 were female, similar to the distribution of the company's employee base.
We observe an imbalance in the use of computers for gender and age: the majority of the branch workers that mostly use one specialized program are female, and are skewed to older ages, while users that work with or without computers are more uniformly distributed.

\subsection{Study Group Sampling}

Following on our advice, the company assigned each of the \numusers participants to one of 12 different \textit{user groups} generated by combining the settings administered for different tested tools and mechanisms:

\begin{itemize}
	\item \textbf{Warnings (3 settings)}: every participant received one of three possible settings on their simulated phishing emails: a simple warning; a detailed warning; or no warning, as a control setting. 
	\item \textbf{Training (2 settings)}: every participant that failed one simulated phishing attack by performing the dangerous action could either be redirected to the training page; or receive no such training, as a control setting.
	\item \textbf{Report feedback (2 settings)}: after reporting a suspicious email, participants could always receive the result of their report as feedback; or they could receive the result only when they reported a legitimate email, as a control setting.
\end{itemize}

For example, Group 1 was administered simple warnings, training, and no feedback after correctly reporting phishing, while Group 2 had the same configuration except receiving complex warnings, and so on. 
Each participant was randomly assigned to one of the $3 \times 2 \times 2 = 12$ groups, so that they were approximately of the same size: from \smallerGroup participants on the smaller to \largerGroup participants on the larger.

\subsection{Experiment Execution}
\label{sec:execution}

From July 2019 to October 2020, the company sent \numEmails different simulated phishing emails to each of the \numusers participants.
The participants received the first 6 emails in random order and at random time intervals~\cite{wash2018provides} during the first 12 months of the experiment (July 2019--July 2020).
They received the last two phishing emails\footnote{The last emails were supposed to be three; however, a simulated \textit{CEO fraud} phishing attack~\cite{mansfield2016imitation} caused unwanted confusion inside the company and
this specific simulated phishing email was canceled.} from August 2020 to October 2020, again in random order and at random time intervals.
Participants were not aware of our study specifically, to not modify their behavior~\cite{parsons2015design,dodge2007phishing}; however, they were aware that the company may occasionally send phishing exercises to their employees.

The \numEmails different email campaigns, of varying difficulty, were designed to simulate broad phishing campaigns targeted to the organization as a whole, rather than sophisticated, individually-crafted spear phishing.
Each different email represented a typical phishing scenario, such as prompts to check their corporate credentials, migrate their email accounts to a new system, or parcel delivery notes as attachments, and used different triggers, such as a sense of authority or urgency or leveraging people's curiosity~\cite{burns2019spear}.
Five emails contained a link to a phishing website, while three had an attached file.
We provide the English version of selected emails in Appendix~\ref{app:emails}. 

During the experiment, our partner company recorded the following interactions with the simulated emails:
\begin{itemize}
	\item \textbf{Clicks} on the links contained in the email;
	\item \textbf{Dangerous actions}: further falling for the phish by, e.g., submitting credentials to the linked website, or enabling macros on the attached document.
\end{itemize}

The company also recorded participants' reports of suspicious emails. 
For each reported email, they stored whether it was one of our simulated phishing emails, the result of the secondary analysis by the anti-phishing appliance, and whether any employee from the IT department looked at such result and confirmed or subverted its verdict.
During the last 5 months of the experiment, the company also recorded how many inbound emails were similar to a reported one in a 20-days window around the date of the report.

At the end of the experiment, we administered a questionnaire with 27 closed-ended questions to 1000 randomly selected participants.
Participants that accepted to respond were informed that their replies were recorded anonymously and would further not be shared with their employer, to encourage honest answers.
The first questions asked participants about knowledge of phishing and other email threats, questions about email warnings, the button to report phishing, contextual training, and whether they recalled falling for phishing.
We report selected questions from the questionnaire in Appendix~\ref{app:questionnaire}.
We received \numquestionnaire complete answers.

\subsection{Ethics and Safety}
\label{sec:ethics}

\paragraph{Study approval} 
This study was initiated and approved by the CISO of our partner company. During the study, we never had access to any PII, and were only given access to anonymized data after collection by the company (see Section~\ref{sec:study_organization}). Since the analysis of anonymized data does not require IRB approval according to our institution's guidelines, we did not submit a formal request. 

\paragraph{Risks to participants}
Our partner company informs its employees about their phishing awareness campaign, that includes phishing exercises. Thus, our study participants were generally aware that the company may send them simulated phishing emails.
The company did not specifically inform participants about the simulated phishing emails that were sent as part of this study (i.e., no informed consent or debriefing).
Participants not in the embedded training group were not specifically informed about the simulated phishing emails, while participants in the embedded training group were informed that the email was phishing if they fell for it.

Our participants were subject to minimal risk as part of this experiment: they were not exposed to greater risk than what they encounter as part of their normal daily life~\cite{finn2007designing}, because they receive real phishing and other malicious emails regularly.
Experiments such as the one we conducted here can have negative impacts such as wasting employees’ time or creating distrust towards the company~\cite{finn2007designing}. This experiment took place as part of the company's existing training program; given this context, we felt that the scientific impact of our experiment merited these potential negative impacts

\paragraph{Data collection and protection}
During the study, our partner company collected data regarding clicks and dangerous actions, and data on emails reported as phishing by participants.
If a study participant entered their password on the simulated phishing web page, our partner company did  not record the entered credentials nor checked if they were correct.
The collected dataset was accessible to a small number of employees working in the IT security department of our partner company and protected with two-factor authentication.

The collected dataset was provided to us in anonymized format such that only attributes like gender, age, and level of computer use were preserved.
Our partner company used the dataset internally to assess its overall exposure to phishing threats, and ensured us that the dataset will not be used for any other purpose, such as employee performance assessment.

The reported emails did not carry any PII: every report recorded whether the reported email was a simulated one or not, and scores and verdicts of the anti-phishing appliance. 
None of these information can link to the original sender, subject, or content of the message.

\subsection{Experiment Statistics}

Overall, the study participants clicked on 6,680 out of 117,864 simulated phishes (5.67\%). During the 15 months, 4,729/14,733 participants (32.10\%) clicked on at least one phish.
The trend for dangerous actions is similar, with the numbers slightly lower: participants fell for 4,885 simulated phishing emails (4.14\% of the total sent emails, and 73.13\% of all the clicked simulated phishes), and 3,747/14,733 participants (25.43\%) users did at least one dangerous action.

There were 4,260 study participants that reported at least one email. In total, the participants reported 14,401 emails, of which 11,035 were our simulated emails.
The button to report phishing was also deployed to 6300 employees that were not part of the experiment but could report phishing: 1,543 of them reported at least one suspicious email, and they reported 4,075 emails.
Thus, the total number of reported emails we received during the 15 months was 18,476.

%% file: sections/results_demo.tex
\begin{figure*}[!t]
	\centering
	\begin{subfigure}[b]{0.32\linewidth}
		\centering
		\includegraphics[width=\linewidth]{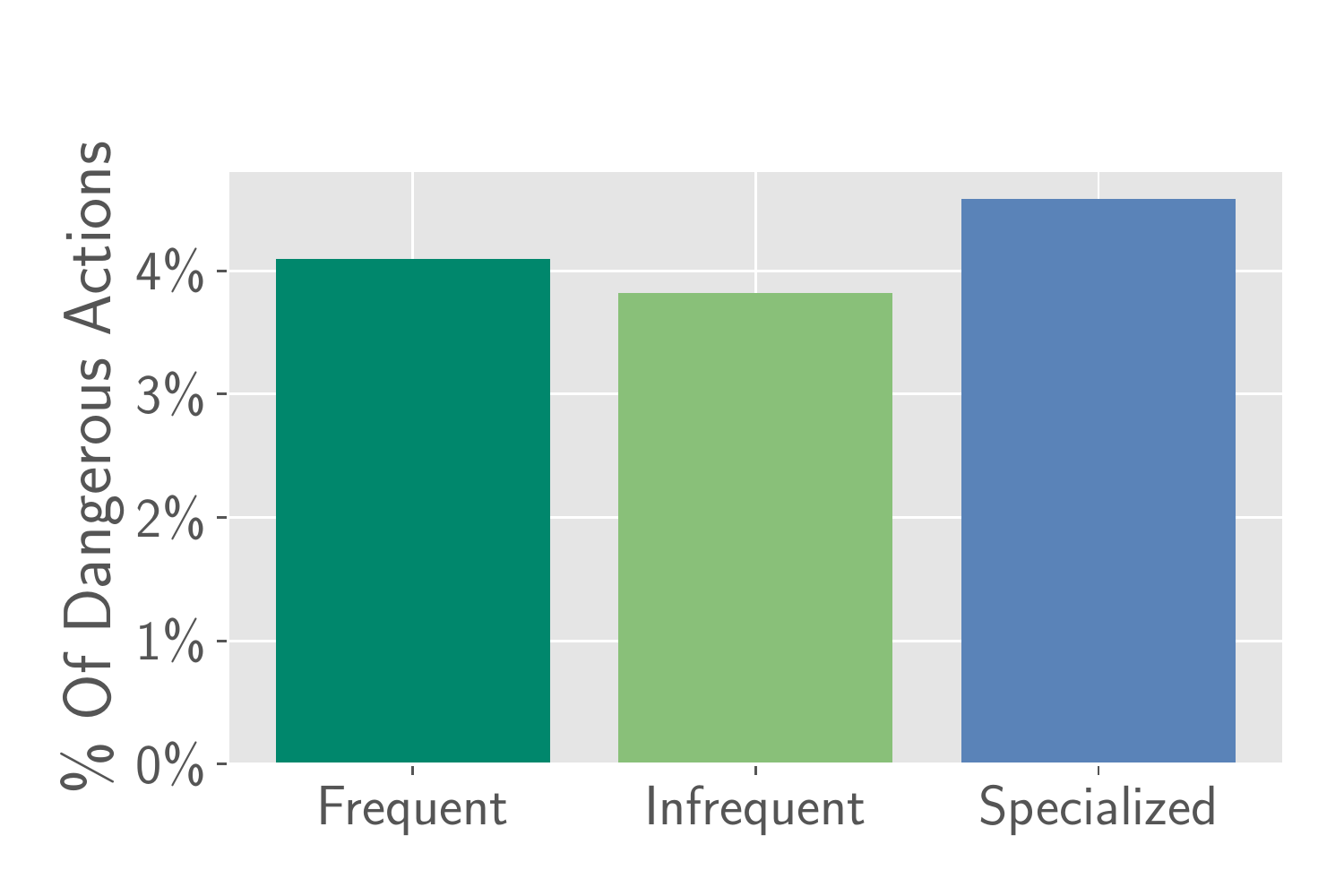}
		\caption{Use of computers in job type.}
		\label{fig:demographics_tech}
	\end{subfigure}
	~
	\begin{subfigure}[b]{0.32\linewidth}
		\centering
		\includegraphics[width=\linewidth]{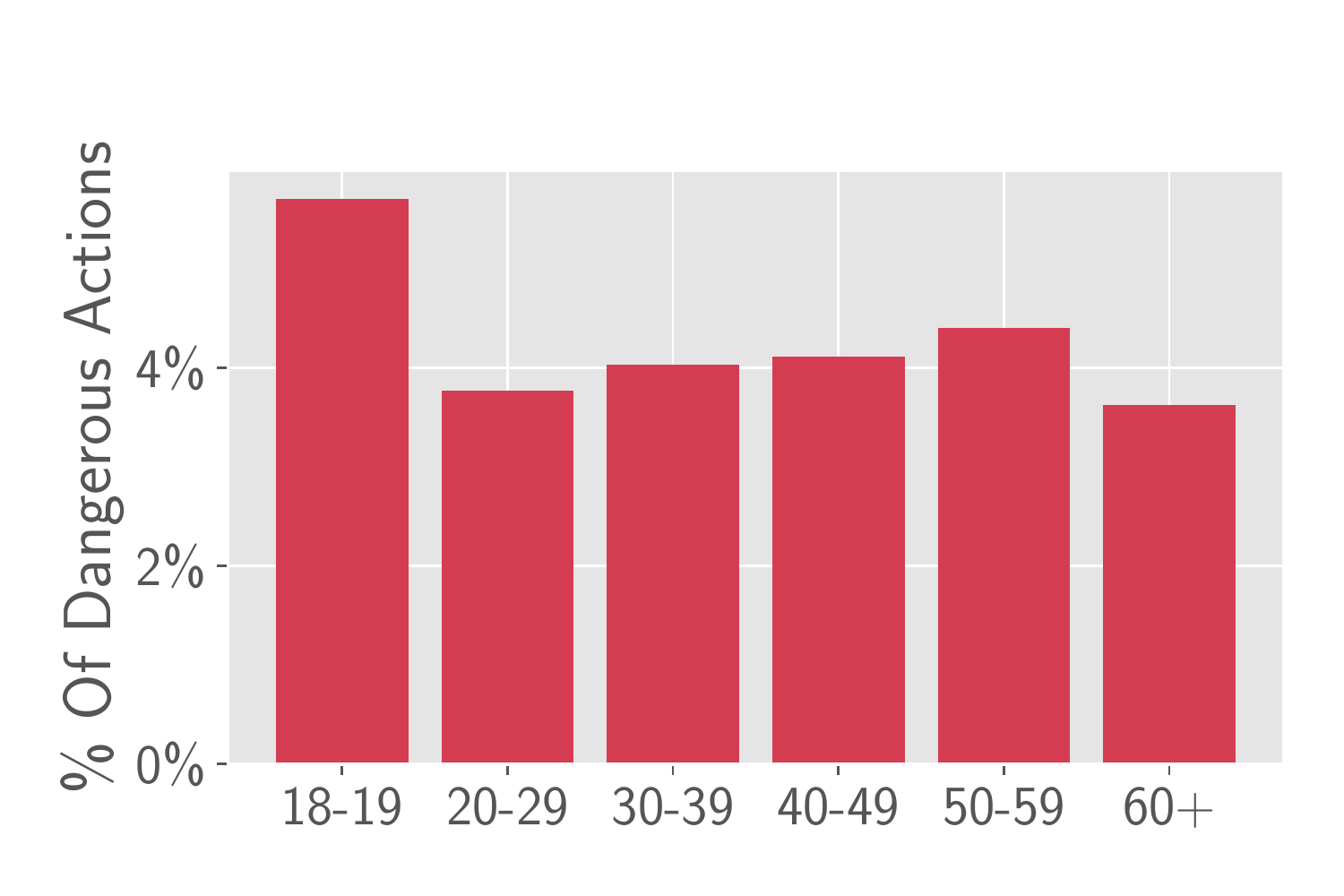}
		\caption{Age Range.}
		\label{fig:demographics_age}
	\end{subfigure}
	~
	\begin{subfigure}[b]{0.32\linewidth}
		\centering
		\includegraphics[width=\linewidth]{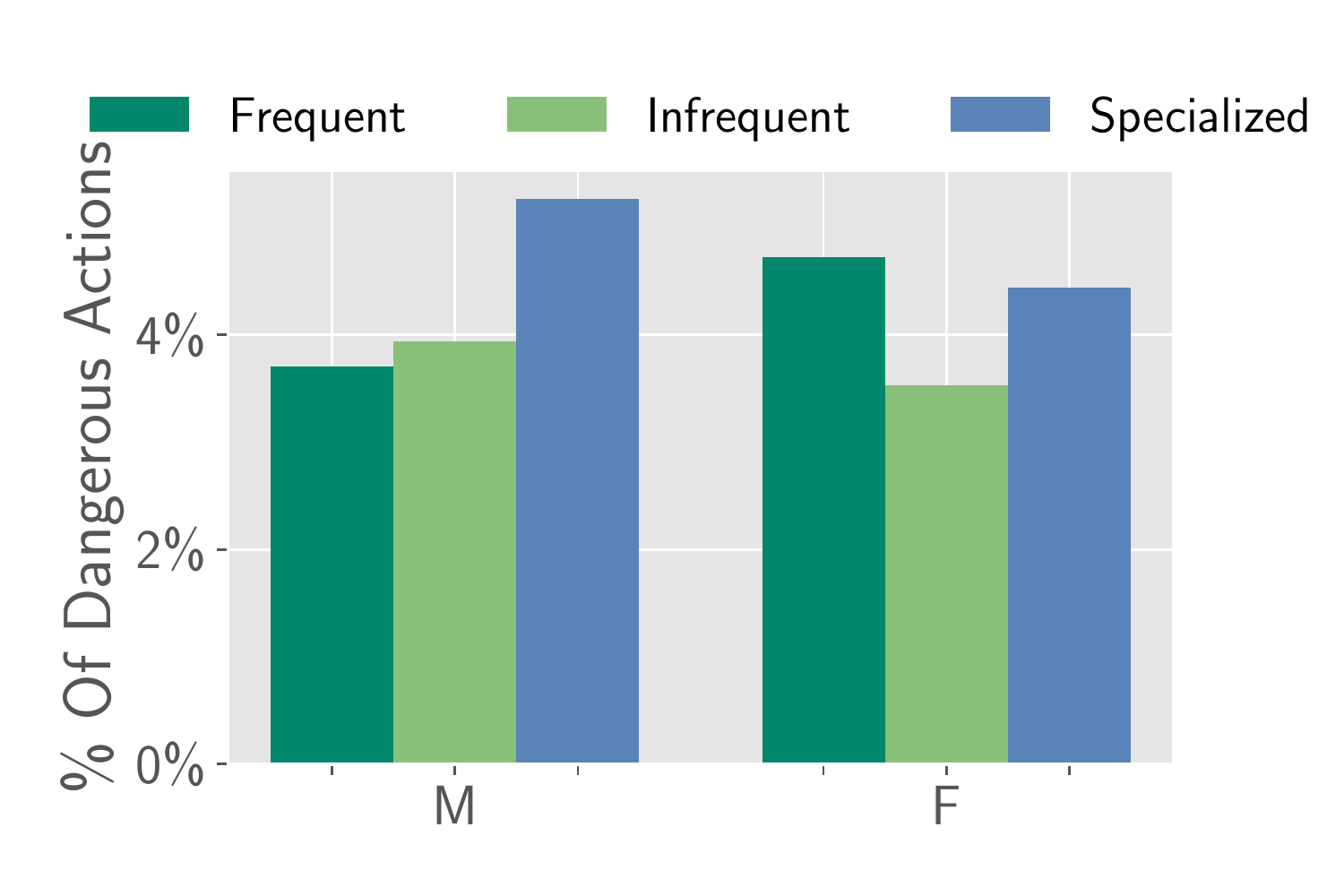}
		\caption{Gender by use of computers in job.}
		\label{fig:demographics_gender}
	\end{subfigure}
	\caption{Percentage of dangerous actions performed out of all phishing emails sent, divided by different demographics. Frequent use of computers but in a very specialized setting, and young and older age all influence the susceptibility to phishing.}
	\label{fig:demographics}
\end{figure*}

In this section, we analyze the experiment data to understand which employees are the most likely to fall for phishing (\RQ).
Recall from Section~\ref{sec:exp_setup} that we classify participants based on Frequent, Infrequent and Specialized use.
We count the number of clicked links and dangerous actions based on demographics and job categories (see
Figure~\ref{fig:demographics}). 
For our following analysis, we define the following three hypotheses:

\begin{itemize}
    \item \texttt{H1}: \textit{Employees' use of computers in their job correlates to falling for phishing}.

    \item \texttt{H2}: \textit{Employees' age correlates to falling for phishing}. 

    \item \texttt{H3}: \textit{Employees' gender correlates to falling for phishing}.
\end{itemize}

To analyze the measured numbers, we fit a linear model with Type III sum of squares to analyze both the demographic properties by themselves, and to capture the interactions among them.
This statistical tool allows us to measure the impact of the independent variables (i.e., the demographic properties) on the dependent variables: number of clicked links and dangerous actions, that we use as proxies for phishing susceptibility.
We fit the model with all the combinations of demographic properties, and exclude the non-significant factors until we obtain a final model with  following results.

\paragraph{The results support \texttt{H1}: correlation with computer use}
As can be seen from Figure~\ref{fig:demographics_tech}, participants whose job type involve Specialized computer use (e.g., branch workers who mostly use a single dedicated program) clicked on more links in phishing emails and performed more dangerous actions than participants in the other comparable groups (Frequent and Infrequent use).
Our fitted model shows that computer use is significant (clicks: $F(2, 14710) = 11.01, p < 0.001$; dangerous actions: $F(2, 14710) = 9.45, p < 0.001$) and a Tukey HSD post-hoc test confirms that the difference between Specialized use and the other two groups is significant both for clicks and dangerous actions. 
However, the difference between Frequent and Infrequent use is not significant.
Thus, while we support previous work that showed relationship between phishing susceptibility and knowledge of technology~\cite{jampen2020don}, this last observation invites to caution, as this relationship seems more nuanced.
While it is common to leverage the \emph{amount} of computer use in participants' jobs as a proxy for technological skills, our results suggest that the \emph{type} of computer use and the expectations in one's job might also influence phishing susceptibility. For example, Specialized use participants in our partner organization may be expected to interact with emails more than Infrequent use participants, who may, therefore, be more suspicious of incoming emails.

\paragraph{The results support \texttt{H2}: correlation with age}
The youngest employees clicked more and performed more dangerous actions.
Our model confirms the interaction between age and phishing susceptibility (click rate $F(5, 14710) = 4.70, p < 0.001$); dangerous action rate ($F(5, 14710) = 3.84, p < 0.001$).
We ran a Tukey HSD test to analyze which groups were more at-risk and confirm what Figure~\ref{fig:demographics_age} shows: participants aged 18--19 were much more likely to click on phishing links and perform the dangerous action than any other age group; participants in the 50--59 age range were also more at risk than the top performers aged 20--29 and 60+. This result supports previous literature~\cite{sheng2010falls,blythe2011f,kumaraguru2009school}.

\paragraph{The results do not support \texttt{H3}: correlation with gender}
Our participants' computer use w.r.t. their gender is not uniform (recall Figure~\ref{fig:setup_users_gender}). Thus, further dividing interactions of both genders by use of computers shows a large difference among the same gender, shown in Figure~\ref{fig:demographics_gender} and confirmed by our model: the combination of gender and computer use is significant (click rate $F(2, 14710) = 13.06, p<0.001$), but gender by itself is not (click rate $F(2, 14710) = 0.23, p=0.63$). 
Indeed, Figure~\ref{fig:demographics_gender} shows us that while Frequent use females were more susceptible than Frequent males, Specialized use males were more susceptible than their female counterpart.
Thus, phishing susceptibility of participants can be better explained by considering the imbalance in job types, contradicting some previous studies~\cite{oliveira2017dissecting,iuga2016baiting}.

%% file: sections/results_time.tex
In this section, we leverage our 15-month study to analyze how the phishing susceptibility of the organization evolves over time (\RQQ).  
To do so, we analyze trends of clicks and dangerous actions over time: how many times (out of maximum 8) participants interacted with the phishes, and how many participants over time eventually did so at least once.

\begin{figure}[tb]
	\centering
	\includegraphics[width=0.8\linewidth]{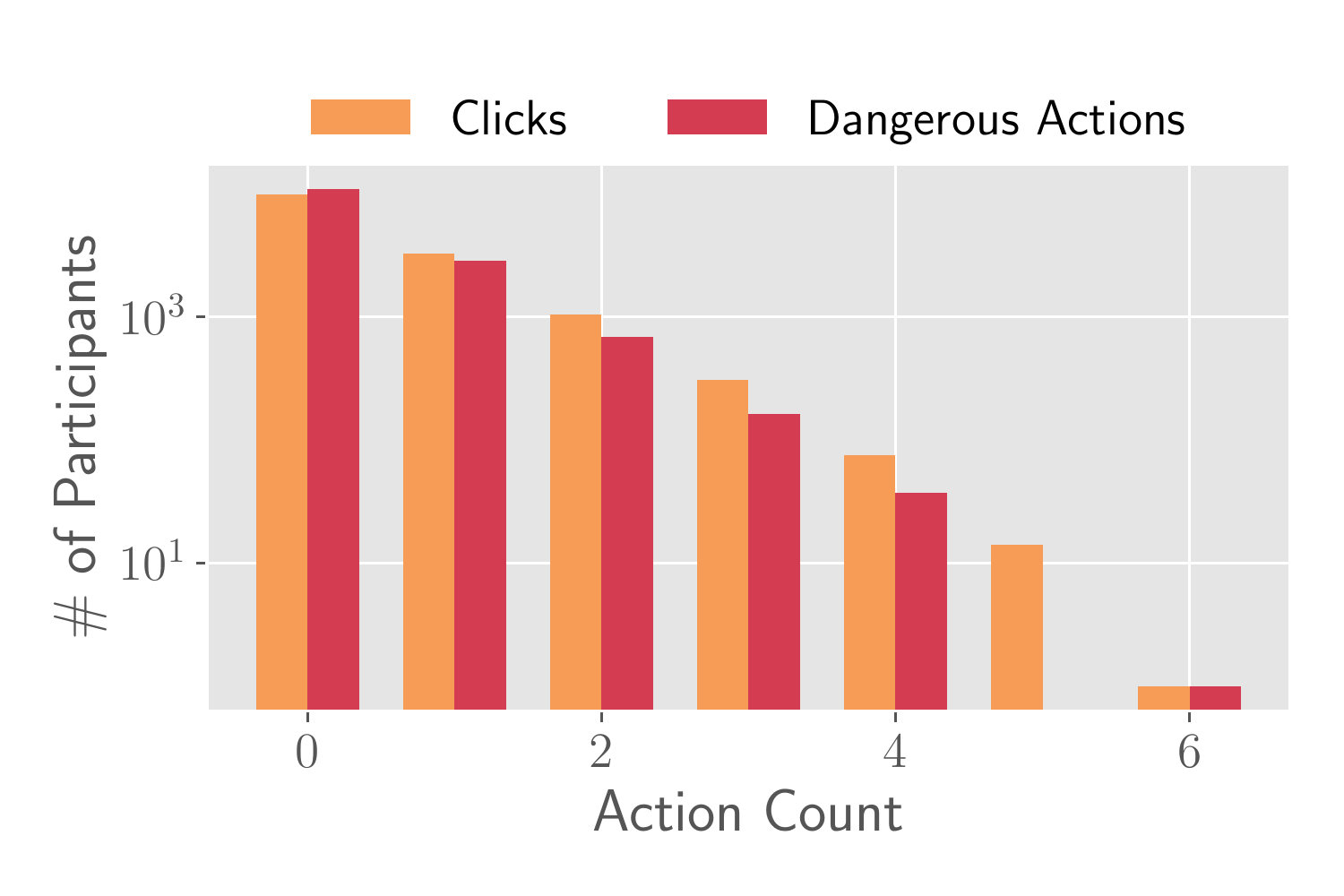}
	\caption{Number of simulated phishing emails that participants clicked or performed the dangerous action (\numEmails being maximum; a missing bar denotes zero participants).}
	\label{fig:repeated_clicks}
\end{figure}

\paragraph{Repeated clickers}
We report in Figure~\ref{fig:repeated_clicks} the histogram of how many participants clicked or performed the dangerous action on the simulations a given amount of times.
A total of 1,448 (30.62\%) participants clicked on two or more phishes, and 896 (23.91\%) performed the dangerous action on two or more---one participant even fell for 6 out of 8 simulations.
Thus, we observe that there will be a small number of employees that will click or fall for phishing emails multiple times, supporting a previous preliminary study~\cite{canham2019enduring}.
Similarly to the raw amount of clicks and dangerous actions, we observe a correlation between age groups and clicking (Welch-corrected ANOVA $F(5, 4199) = 5.72, p<0.001$) or performing the dangerous action ($F(5, 4186) = 3.66, p = 0.002$) on more than one simulated phishing emails. In both cases, a Tukey HSD test shows that the younger group of participants aged 18-19 stands out as the one more likely to click more than once.

\paragraph{Many employees will eventually fall for phishing if continuously exposed}
In our experiment 4,729 out of 14,733 (32.10\%) participants clicked on at least one link or attachment in our simulated phishing emails.
A similar high number applies to dangerous actions: 3,747 out of 14,733 (25.43\%) performed at least one. These results indicate that a rather large fraction of the entire employee base will be vulnerable to phishing when exposed to phishing emails for a sufficiently long time. 
We are the first to show such result at scale.

%% file: sections/results_tools.tex
In this section we analyze the data collected from our experiment to answer \RQQQ related to the effectiveness of phishing warnings and training.

\subsection{Effectiveness of Warnings}
\label{sec:warnings}

Recall from Section~\ref{sec:exp_setup} that we experimented with two types of warnings (short and detailed), and a control group that did not see any warnings. To analyze the effectiveness of these two warning types, we use the following hypotheses:

\begin{itemize}
    \item \texttt{H4}: \textit{Adding warnings on top of suspicious emails helps users in detecting phishing}. 

    \item \texttt{H5}: \textit{Detailed warnings are more effective than short ones}.
\end{itemize}

\begin{figure}[t]
	\centering
	\includegraphics[width=0.8\linewidth]{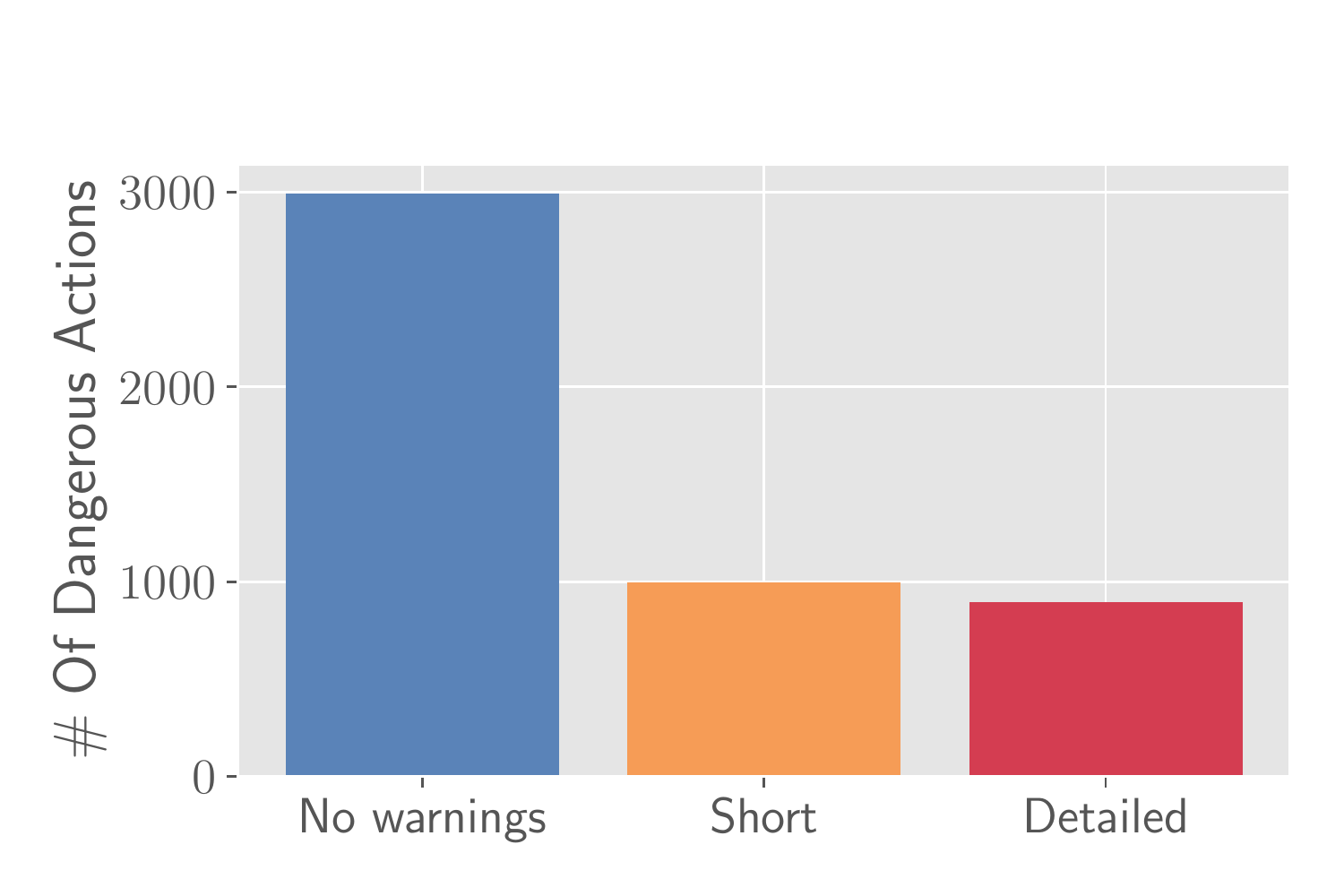}
	\caption{Dangerous actions by administered warning. Both warning types helped the participants significantly.}
	\label{fig:snippets_fall}
\end{figure}

\begin{figure}[t]
	\centering
	\includegraphics[width=0.8\linewidth]{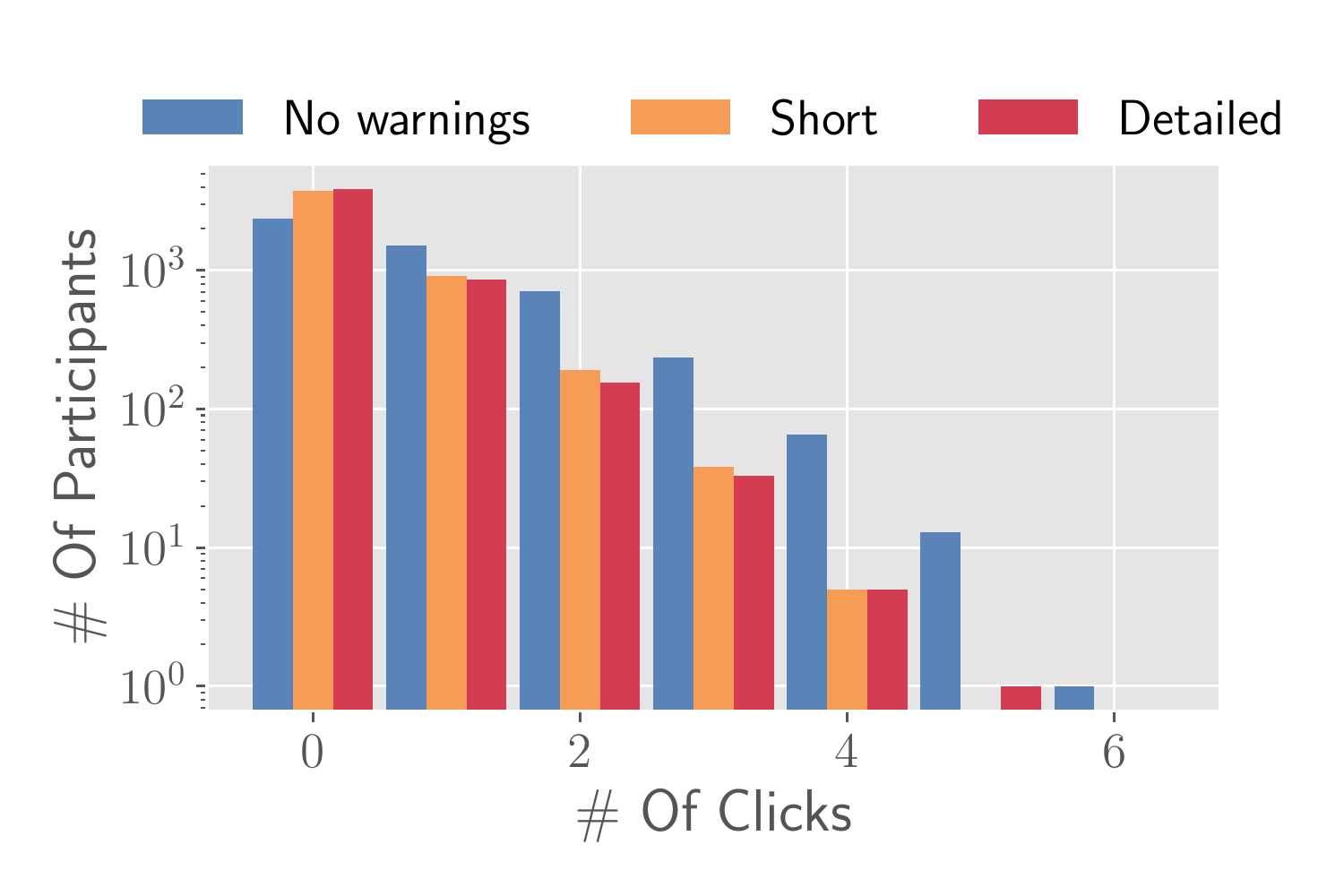}
	\caption{Number of different phishing emails that participants clicked on, by administered warning. Missing bars denote a 0.}
	\label{fig:snippets_histogram}
\end{figure}

\paragraph{The results support \texttt{H4}: warnings help users}
Figure~\ref{fig:snippets_fall} shows click and dangerous action rate for the different warning configurations.
We observe that both types of warnings greatly helped participants both in avoiding clicking on links in our simulated phishing emails and not falling for the phish by performing the dangerous action.
Considering click rate, the group with no warnings clicked 3,964 times, compared to the lower 1,427 clicks for short and 1,289 clicks for long warnings (Welch-corrected ANOVA $F(2, 7485) = 564.71, p<0.001$).
Dangerous actions rate is similar: 2,994 dangerous actions for no warnings, compared to 998 and 893 dangerous actions, respectively ($F(2, 7461) = 392.58, p<0.001$).
Figure~\ref{fig:snippets_histogram} shows the histogram of how many participants clicked on a simulated phish a given amount of times.
We observe a strong correlation between receiving any of the warnings and not clicking or performing the dangerous action more than once (clicks: $F(2, 9287) = 358.88, p < 0.001$, dangerous actions: $F(2, 9194) = 239.68, p < 0.001$).
Our results support this widespread industry practice~\cite{gmailwarnings}.

\paragraph{The results do not support H5: detailed warnings are not more effective than short ones}
To check whether there is any difference between short and detailed warnings, we ran a Tukey HSD test between all groups and observed that, while both warnings correlate with lower total clicks and dangerous actions, there is no significant difference between short and detailed warnings.
Thus, the way we provided additional information to users (by mimicking the current industry practices, rather than making radical changes to email warnings~\cite{petelka2019put}) does not seem to provide better phishing protection.

\subsection{Effectiveness of Contextual Training}
\label{sec:contextual_training}

Recall from Section~\ref{sec:exp_setup} that we tested the effectiveness of contextual training after falling for a simulated phishing email by administering it only to half of the participants---the other half was a control group for training and did not see the webpage.
We formulate the following hypothesis:
\begin{itemize}
	\item \texttt{H6}: \textit{Receiving contextual training helps users improve in future phishing detection}. 
\end{itemize}

We analyze both the frequency at which participants clicked or performed the dangerous action, and the correlation between training and doing more than one click or dangerous action.

\begin{figure}[t]
	\centering
	\includegraphics[width=0.8\linewidth]{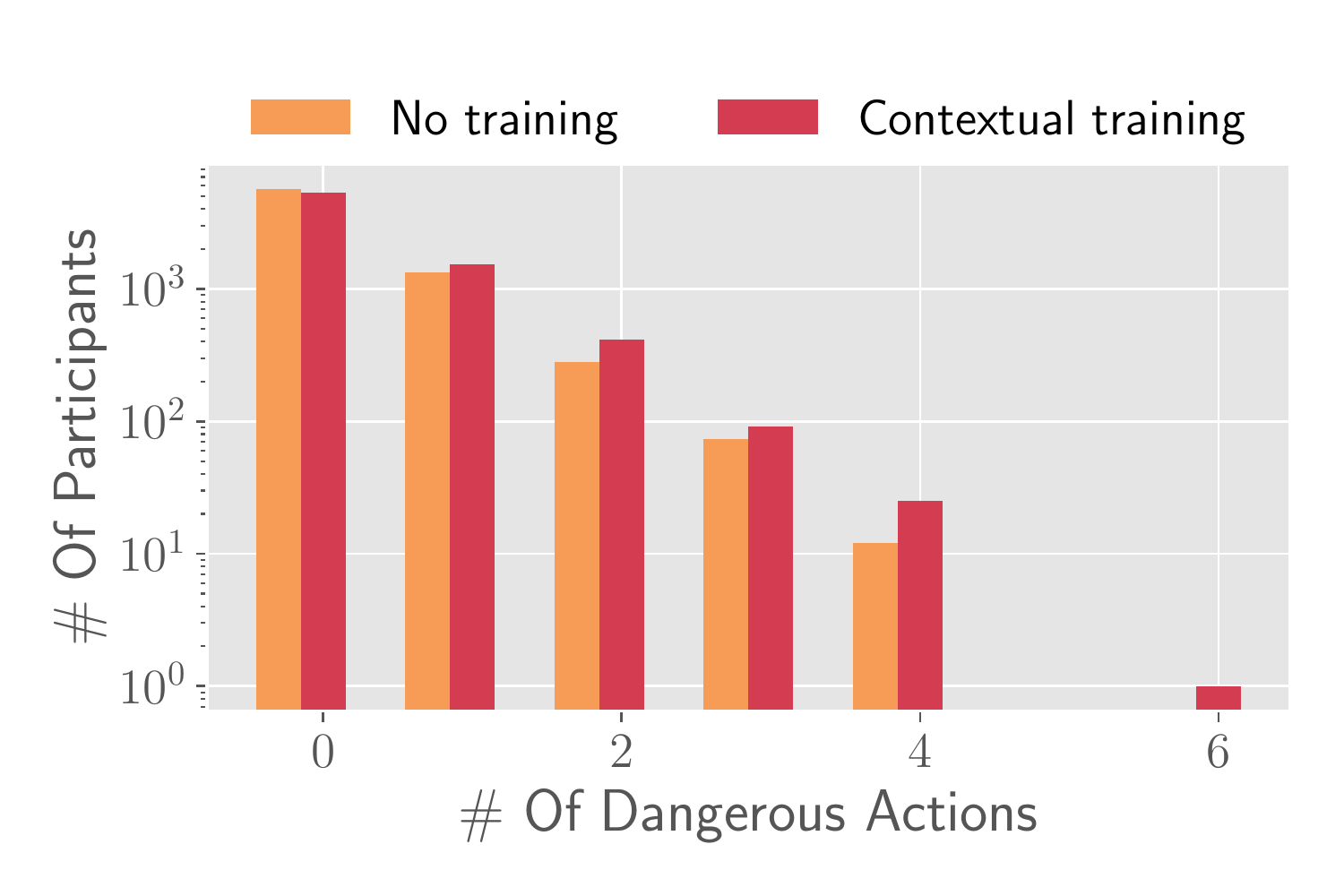}
	\caption{Number of different simulated phishing emails participants performed the dangerous action on, by administered contextual training. Missing bars denote a 0.}
	\label{fig:training_histogram}
\end{figure}

\paragraph{The results do not support \texttt{H6}: voluntary contextual training does not improve future phishing detection}
Surprisingly, we observe that both click and dangerous actions rates are higher for participants that received contextual training (i.e., participants who were forwarded to a training page) after falling for simulated phishes: for clicks, 3,087 versus 3,593; for dangerous actions, 2,155 versus 2,730.
Figure~\ref{fig:training_histogram} shows the histogram of how many different phishing emails participants performed the dangerous action on.
As expected, the number of participants that did not fall for any simulated phish, or fell only once, is similar among the two groups: such participants that were in the training group either never saw the training page, or saw it after performing their only dangerous action.
However, if we focus on participants that fell two or more times (and thus, on participants in the training group that fell again for phishing after being shown the training page), we see that the distribution is more skewed to the right for participants in the training group.
Indeed, participants that clicked on two or more phishing emails were 647 without training, and 801 with training.
This shows a strong correlation between the provided training page and clicking on phishing emails or even performing the dangerous action more than once (Welch-corrected ANOVA for clicking: $F(1, 14592) = 18.37, p < 0.001$; dangerous actions: $F(1, 14279) = 33.80, p < 0.001$).

This perhaps surprising result requires a careful interpretation. 
What our experiment showed is that \textit{this particular way} of delivering voluntary training 
does not work. Instead, such training method may cause unexpected and negative side effects, such as increased susceptibility to phishing.
This finding is significant, because the tested phishing training delivery method is a common industry practice~\cite{rapid7,wombattraining,knowbe4,cofense}, and the training material (refer to Section~\ref{sec:exp_setup}) was designed by a specialized company according to known guidelines and best practices from previous work~\cite{kumaraguru2007getting,kumaraguru2009school,kumaraguru2010teaching}.
It would be interesting to study whether other possible ways to deliver contextual training (e.g., ones where interaction with the provided training material is enforced) would work better.
Our study did not test the effectiveness of mandatory training.

To gain some insights on why susceptibility to phishing \emph{increased} among those participants who were forwarded to the training page, we analyzed the answers to our post-experiment questionnaire.
One possible explanation that emerges from the questionnaire responses was a false sense of security that is related to the deployed training method: out of the respondents who remembered seeing the training page, 43\% selected the option \emph{``seeing the training web page made me feel safe''}, and 40\% selected the option \emph{``the company is protecting me from bad emails''}.
It remains an open question for future work to explore whether this is due to a misinterpretation of the training page (i.e., whether the participants thought they were protected from a real attack), or if this is because of overconfidence in the organization's IT measures in general, as observed in similar settings in the past~\cite{williams2018exploring,greene2018user,conway2017qualitative}. 

Ultimately, our result shows that organizations need to be careful when using this training method, and aware of possible unintended side effects.

%% file: sections/results_ucp2.tex
We now analyze the data collected from our experiment to answer \RQQQQ, related to crowd-sourced phishing detection in an organization.
Such method needs to fulfill the following requirements to be useful:

\begin{itemize}
	\item \textbf{Sustainability}: employees need to keep reporting suspicious emails over long period of time.
	\item \textbf{Effectiveness}: employees' reports need to be sufficiently accurate and timely so that the organization can stop new campaigns quickly enough.
	\item \textbf{Practicality}: the operational workload to process all the reported emails needs to remain acceptable.
\end{itemize}

\subsection{Reporting Sustainability for Employees}
\label{sec:crowd_sustainability}

Recall from Section~\ref{sec:exp_setup} that we decided to experimented with two types of feedback: (i) always receive the result of their report; or (ii) receive the result only when (erroneously) reporting a legitimate (non-phishing) email.
To investigate reporting sustainability, we examine how the employees' activity in reporting suspicious emails evolved over time, and whether the tested method of encouraging reporting worked.
We examine these questions using the following two hypotheses:

\begin{itemize}
	\item \texttt{H7}: \textit{Employees keep reporting over time at a steady rate}
	\item \texttt{H8}: \textit{Providing feedback to reports encourages to report again in the future}
\end{itemize}

We count all reports and analyze their rate over time, and compare the number of participants that reported more emails after receiving the two different types of feedback.

\begin{figure}
	\centering
	\includegraphics[width=0.8\linewidth]{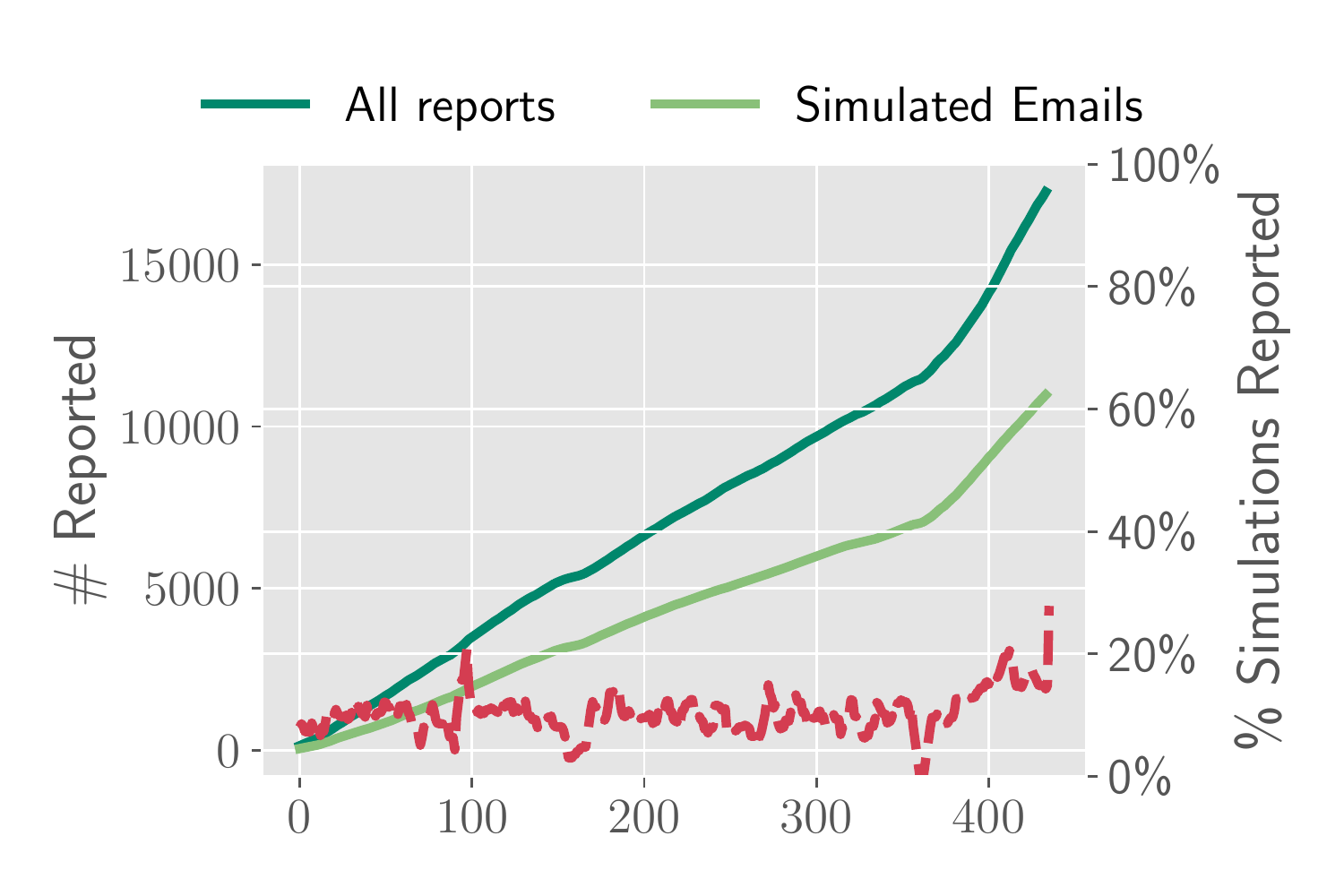}
	\caption{Cumulative email reports over time. The dashed red line shows the percentage of simulated emails reported daily.}
	\label{fig:reports_over_time}
\end{figure}

\begin{figure}
	\centering
	\includegraphics[width=0.8\linewidth]{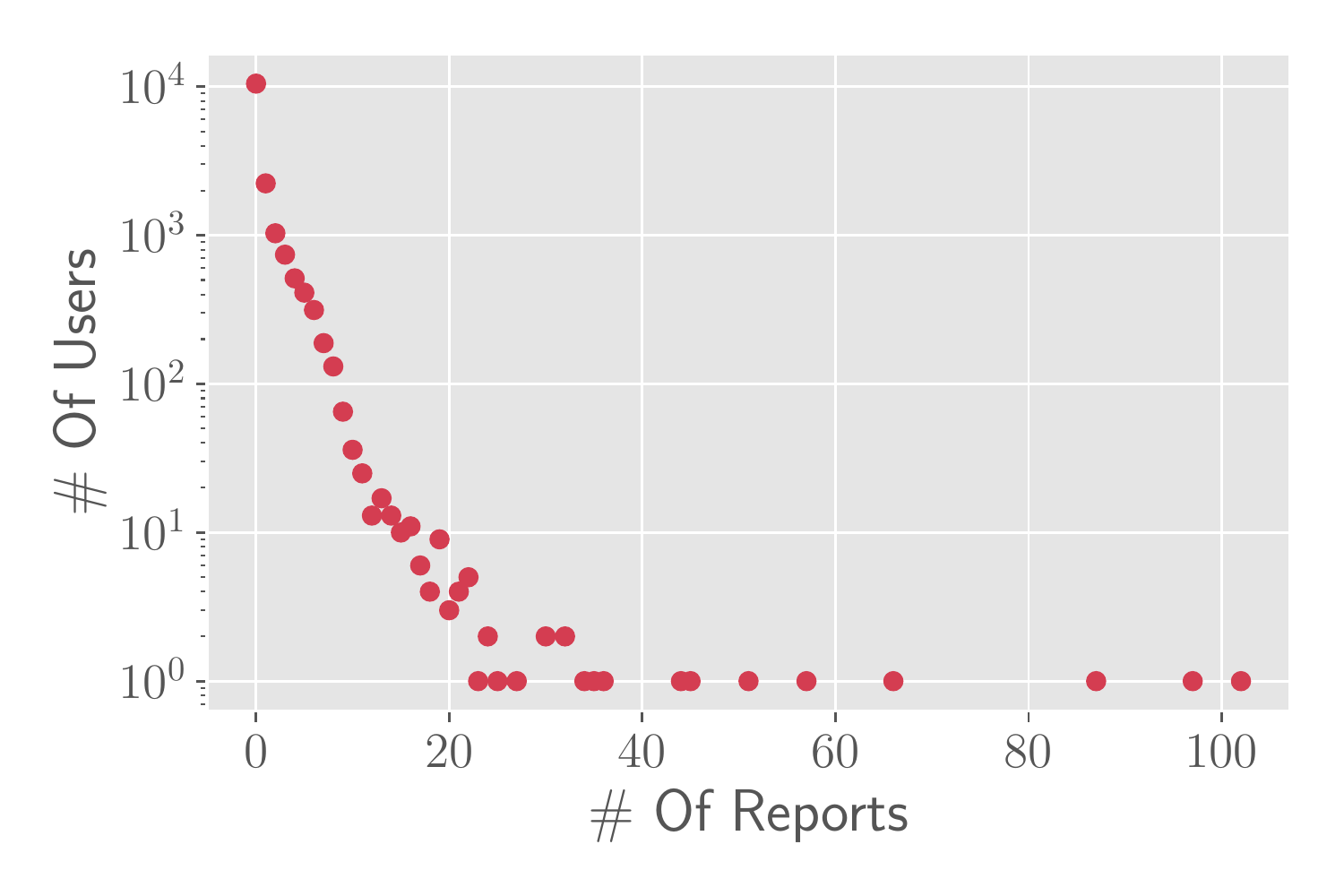}
	\caption{Distribution of the number of reports per user. }
	\label{fig:frequency_total_reports}
\end{figure}

\paragraph{The results support \texttt{H7}: employees continue reporting emails}
Figure~\ref{fig:reports_over_time} shows the number of suspicious emails reported over the duration of the entire experiment.\footnote{These numbers include all \numbuttons employees that received the button.}
We observe a steady income of reports that does not slow down (and even increased when the two new phishing emails were released in August 2020), as shown by the constant fraction of simulated emails reported daily.
We further analyze the distribution of frequency of reports that is shown in Figure~\ref{fig:frequency_total_reports}.
While 90\% of the employees that reported suspicious emails reported 6 or less, there is a non-negligible amount of very active users. We conclude that in our experiment of 15 months, there was no significant ``reporting fatigue'' suggesting that, if reporting is made easy, employees can actively keep on reporting suspicious emails for long periods of time.

Additionally, we examined whether any demographic influences the quantity of reports by fitting a linear model with Type III sum of squares.
Similarly to phishing susceptibility, the combinations of age and computer use in job, and gender and computer use in job are significant (age and computer use $F(10,14710)=6.49, p<0.001$; gender and computer use $F(2,14710)=11.35, p<0.001$).
Considering the skewed distribution of computer use, we assume it is the main contributing factor.
Indeed, we find that Frequent computer use participants reported a very encouraging 22\% of all the simulated emails that they received, while Infrequent use participants reported only 10.20\% and Specialized use 7.60\%. We conclude that, quite intuitively, employees with the best expected computer skills are also the most active reporters. However, interestingly, Infrequent use participants were more active than the Specialized ones.

\paragraph{The results support \texttt{H8}: positive feedback encourages employees to report more}
We find a significant interaction between the type of administered feedback type and the amount of reported emails.
To measure this, we first exclude all the participants that never reported any email.
Then we count how many emails were reported by the group that actually received positive feedback and by the one that only received feedback about false reports.
The former (2,046 participants) is composed by participants in groups that always received feedback and that reported at least one malicious or simulated email (thus receiving the positive feedback).
The latter (2,201 participants) is formed by those in groups that did not receive positive feedback, and by those in a group that could receive positive feedback but only reported legitimate emails (thus never receiving the positive feedback).
We ran a Welch-corrected ANOVA ($F(1, 3224) = 31.62, p < 0.001$) confirming that participants that saw the positive feedback were more likely to report more emails.

\subsection{Effectiveness of Crowd-Sourced Phishing Detection}
\label{sec:reporting-effiency}

To analyze the effectiveness of crowd-sourced phishing detection mechanism as a whole, we analyze \emph{timeliness} and \emph{accuracy} of reports. In addition to sufficiently high reporting activity, organizations need both quick and sufficiently accurate reporting to be able to detect and stop novel phishing campaigns that are often  short-lived~\cite{oest2020sunrise}. 

We note that since we did not send thousands of copies of the same phishing email at the same time, we cannot directly measure how fast such mass phishing campaigns are reported and thus detected. 
Instead, we measured how fast our randomly timed simulated phishing emails were reported by participants. Based on these numbers, we can then estimate how quickly and accurately real mass campaigns could be detected.

\begin{figure}[t]
	\centering
	\includegraphics[width=0.8\linewidth]{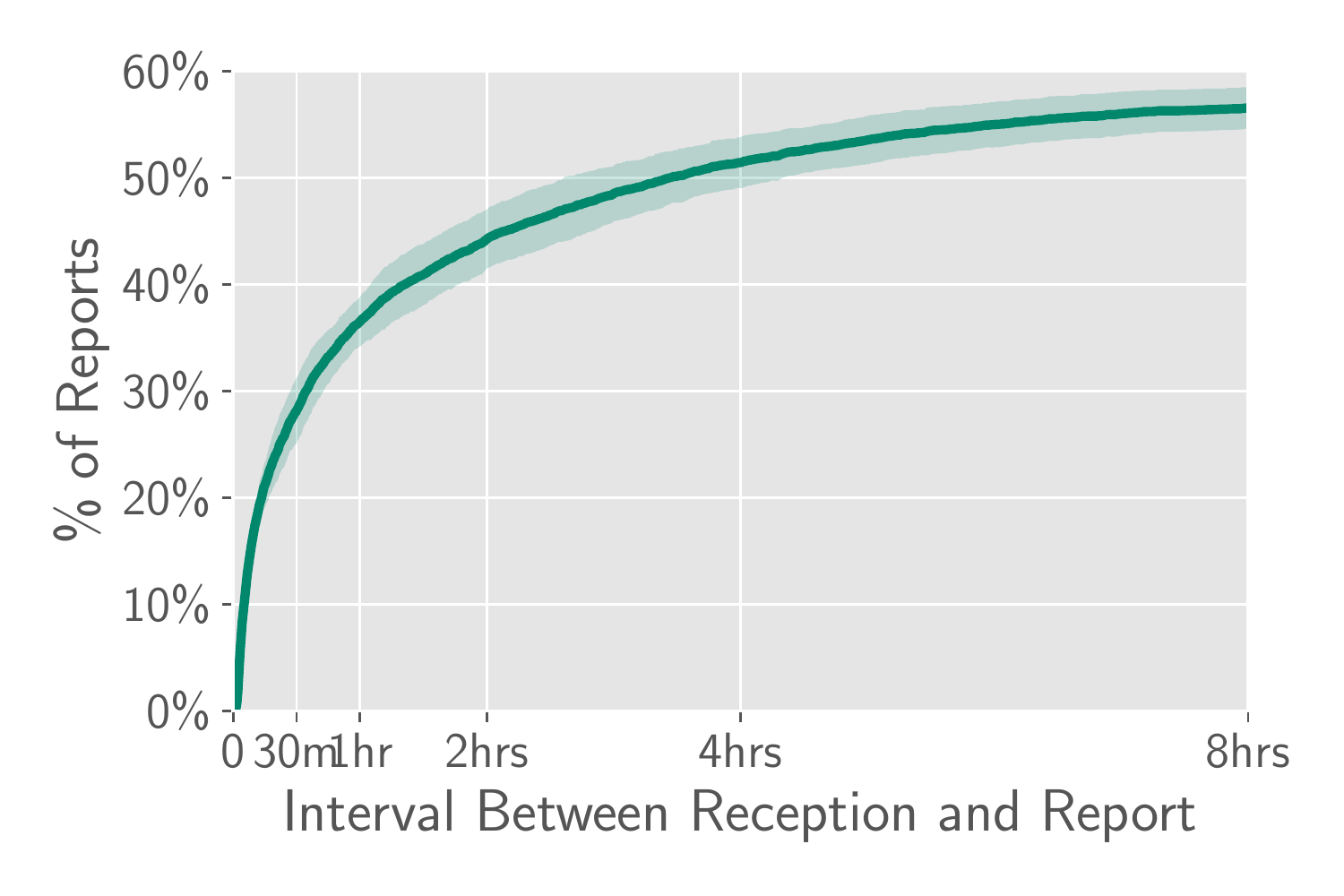}
	\caption{Average per--user--group Cumulative Distribution Function of reported simulated emails as a function of the interval between email reception and report, with standard deviation as a colored region.}
	\label{fig:reports_delta_times}
	\vspace{-.7em}
\end{figure}

\paragraph{Timeliness} 
We show in Figure~\ref{fig:reports_delta_times} the percentage of reports of our simulated emails that arrived shortly after their delivery.
We can observe that the reaction time of the employee base as a whole is fast: on average around 10\% of the reports arrived within 5 minutes; ~20\% within 15; and 30\% to 40\% within 30 minutes. 
We observe no significant difference between the reporting times of different simulated email campaigns: despite all having different number of total reports (from 2,538 reports of the most reported simulation, down to 832 reports for the least reported), all consistently see a similar amount of reports incoming within the first 30 minutes.

To apply these numbers to a hypothetical company of 1,000 employees where 100 of them are targeted by a phishing campaign, we would have between 8 and 25 reports of the email by employees---of which one within 5 minutes with high probability, and a larger number within 30 minutes.

\paragraph{Accuracy}
The average accuracy of reports was good: 68\%, up to 79\% if spam emails should be reported as well.\footnote{As ground truth, we consider here the outcome of the secondary antiphishing appliance, corrected and validated by the IT department of the company.}
We observe that the distribution of employees' accuracy in reporting is wide: while over 60\% of the reporting employees have an accuracy of 80\% or more, there is a non-trivial fraction that was always wrong (13\% if spam should be reported; 22\% otherwise)---however, it mostly comprises employees who reported only a single email.
The accuracy of the top 10\% of very active employees that reported 6 or more emails (recall Section~\ref{sec:crowd_sustainability}) is around 5\% higher than considering all employees.
We further note that very high reporting accuracy is not crucial.
If using a secondary anti-phishing appliance to triage reports, as done in our experiment, employees can be encouraged to be overly-cautious and report emails when in doubt (not only when absolutely sure), as the appliance can serve as a first check on the email, and keep the operational workload acceptable, as discussed below.

\paragraph{Incident awareness}
Additionally, we analyzed the employees' awareness of being the victims of a security incident.
We start by noticing that 6\% of the participants who performed the dangerous action on our simulated emails immediately reported the email, thus realizing they were victims of a phishing attack.\footnote{We only measure participants that did not receive training after falling for a simulated phish, because they had to understand by themselves what happened---the training material stated clearly that it was a simulated phishing attack.}
Only 3.7\% of these participants did not toggle the checkbox of the report button that allowed employees to report whether they visited the link contained in the email, or opened its attachments.
Interestingly, we observe that some participants were overly-cautious: 13\% of the reports of simulated phishing emails stated that they opened the link or attachment, despite not having done it.

\subsection{Practicality for the Organization}
\label{sec:reporting-practicality}

We observe that a secondary appliance triaging the reports makes the added workload reasonable: out of the 7,191 non-simulated reported emails in 15 months, only 689 (9\%) of the decisions were taken by human administrators, and actually overturned the decision taken by the appliance only 50 times (7\% of the total handled cases). 
The main goal of this secondary appliance is to filter out the reports of clear benign emails or minor threats such as spam, which include the majority of our collected reports: out of 7,191 reports of emails not part of our exercise, 3,531 were benign, and 2,371 were spam or unwanted newsletters.
Thus, only roughly 1.5 emails per day needed manual handling from the IT department---a clearly acceptable workload for a large organization that was collecting reports from over 21,000 users.

\subsection{Finding Real Phishing Campaings} 

We further validate our crowd-sourced phishing detection approach by analyzing whether we caught any \emph{real} phishing campaigns delivered to employees of the company (in addition to our simulated phishing emails).
We use the verdicts of the secondary filter and manual inspection by IT specialists to find reported phishing and other malicious emails. 
We observed 918 reports of real phishing emails during the last 5 months of our deployment.
With email similarity techniques, we measured how many emails similar to the reported ones were incoming and found 252 large-scale phishing campaigns comprising 28,830 emails, and 1,534 emails with malware attached that our crowd-sourced approach would have detected in a short time span from their beginning.

%% file: sections/discussion_limitations.tex
\paragraph{Simulated emails limitations}
Recall that 3 of our 8 emails had a malicious attachment where the dangerous action was to enable macros.
While the company could monitor when macros were enabled, with a network call to the monitoring infrastructure, it could not know when participants only \textit{clicked}, i.e., simply opened and closed the attachment without enabling macros.
Thus, for attachments we underestimate the number of clicks by setting it to the number of dangerous actions.

The company did not record when a simulated email was opened, thus we do not know the conversion rate from opening the email to clicking and dangerous actions.

Due to data protection concerns, it was not recorded whether employees submitted their valid credentials to the simulated phishing websites.
Therefore, the amount of employees that we recorded performing some of the dangerous actions (e.g., submitting credentials) could be overestimated, because we cannot filter out employees who submitted bogus credentials.

\paragraph{Email warnings limitations}
Our partner company added warnings on top of simulated phishing emails, but not on top of emails that the inline filtering solution in use deemed suspicious but let through anyway.
This could lead some participants to fall once for our simulated phishing email, and subsequently associate the presence of warnings to surely suspicious emails, or even worse, to training exercises.
Further studies on this promising type of warnings when added on top of legitimate but suspicious-looking emails are needed, to remove the potential bias.

\begin{figure}[t]
	\centering
	\includegraphics[width=0.8\linewidth]{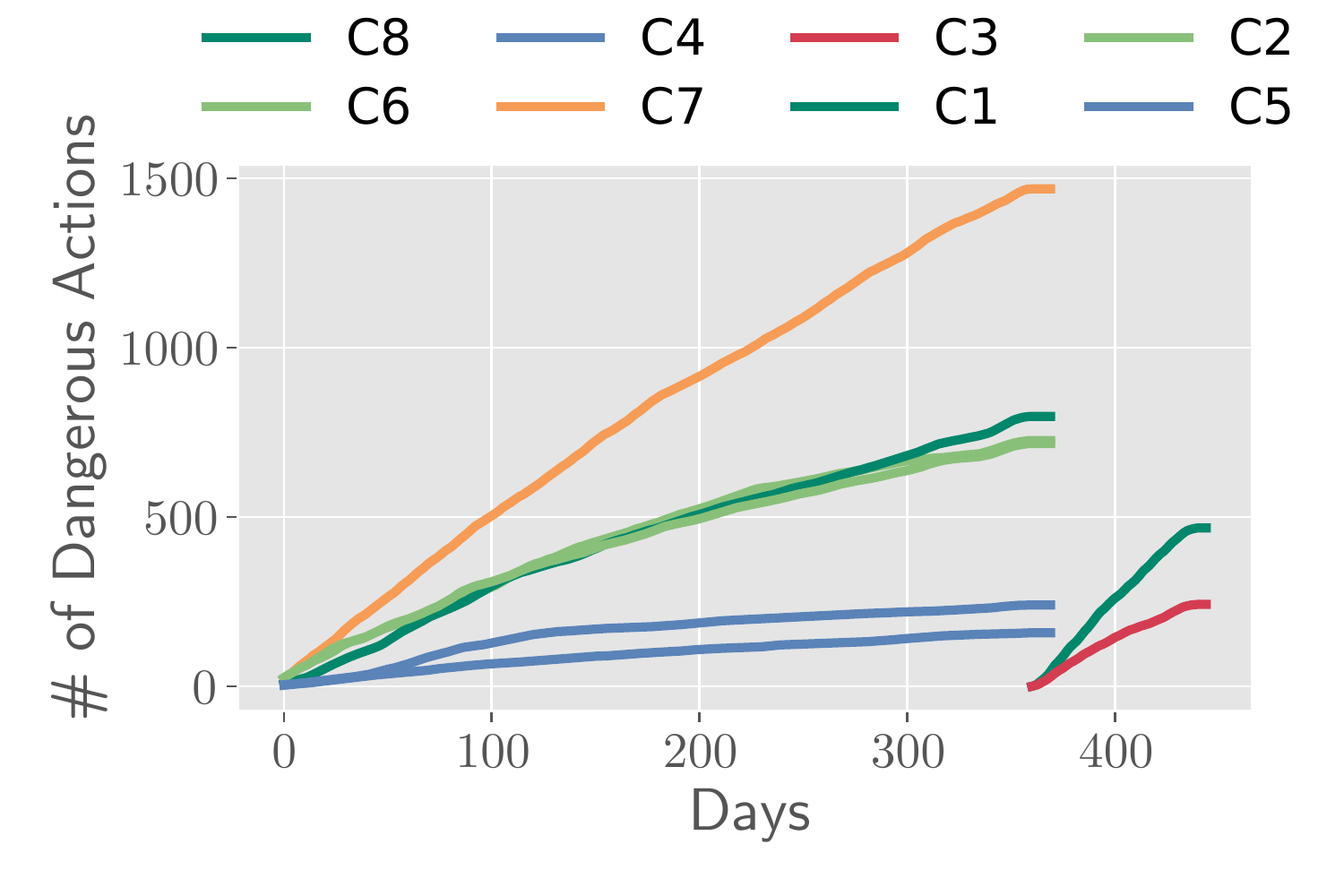}
	\caption{Cumulative count of dangerous actions per campaign during the experiment duration.}
	\label{fig:campaigns_fall_over_time}
\end{figure}

\paragraph{Campaigns success rates}
As shown in Figure~\ref{fig:campaigns_fall_over_time}, the different simulated phishing email campaigns had different success rates. Such differences do not influence our analysis of RQ1-RQ4 due to two reasons. First, we always count the total number of clicks or dangerous actions for all campaigns and compare total counts. Second, the order of administered campaigns was randomized for each participant in large groups.

\paragraph{Applicability to different companies}
Our partner company operates in numerous different sectors, has a diverse workforce, and large size. Thus, we believe that our results can generalize to various similar-sized (large) companies. 
It is unclear whether our results generalize to companies with very specialized IT workers, e.g., software engineering companies, or to very small organizations.

%% file: sections/related_work.tex
\paragraph{Phishing and demographics}
Age is one of the most analyzed factors of phishing, as it intuitively often correlates with technological skills. Studies showed that very young~\cite{sheng2010falls,blythe2011f,kumaraguru2009school}, and older people~\cite{oliveira2017dissecting,lin2019susceptibility,gavett2017phishing} are more at-risk for phishing. 
Preliminary studies show that aging increases susceptibility to phishing~\cite{lin2019susceptibility}, but only the two extremes (very young and senior persons) were tested, not the full age spectrum.
Further, different age ranges are susceptible to different types of phishing emails~\cite{wang2012research,oliveira2017dissecting}.
Gender is a more divisive demographic, according to a recent literature survey~\cite{jampen2020don}, but the studies that do find an impact show that women are more vulnerable~\cite{oliveira2017dissecting}, and can detect less phishing attempts~\cite{iuga2016baiting}.
Experience with computers~\cite{jampen2020don}, experience of previous phishing attempts~\cite{gavett2017phishing}, and seniority at an organization~\cite{burda2020testing} also positively influence phishing immunity.

\paragraph{Phishing at workplace}
Some previous studies show that within the organization's boundaries, employees feel safer and generally trust their company's measures, thus lowering their attention~\cite{williams2018exploring,greene2018user}, 
and the existence of ``repeated clickers'' that are extremely at risk of being phished~\cite{canham2019enduring}. Other studies find that helping employees in phishing prevention is made hard by the fact they struggle to comply with corporate security policies and often ignore them~\cite{tsai2016understanding,siponen2014employees}.

\paragraph{Phishing training and warnings}
There is broad consensus that training should be active, e.g., with security games~\cite{sheng2007anti}.
A popular mechanism is running simulated phishing exercises, an approach adopted by several companies~\cite{rapid7,cofense,wombattraining,knowbe4} following promising research results~\cite{kumaraguru2007protecting,kumaraguru2008lessons,kumaraguru2009school}, where (possibly unaware~\cite{parsons2015design,dodge2007phishing}) employees receive simulated phishing emails over time, ideally at random intervals~\cite{wash2018provides}. 
This practice is often combined with embedded training: immediately redirecting the employee that fell for phishing to a dedicated web page explaining the simulated attack they just fell for and providing information about phishing~\cite{kumaraguru2009school}.

Previous studies suggest that training should be continuous, as knowledge retention spans from a few days~\cite{kumaraguru2008lessons,kumaraguru2010teaching} to a few months at most~\cite{kumaraguru2009school}.
However, research effort has unclear external validity, because most work employed small populations~\cite{kumaraguru2008lessons,kumaraguru2007getting,carella2017impact,wash2018provides,kumaraguru2007protecting,burns2019spear}, was shorter in time~\cite{wash2018provides,kumaraguru2008lessons,burns2019spear}, had populations with little diversity~\cite{kumaraguru2007getting,carella2017impact,burns2019spear} or tested a role-playing setting only~\cite{kumaraguru2007protecting,kumaraguru2007getting}---and a recent study questioned whether these results transfer to a corporate setting~\cite{caputo2013going}.
Further, the business ecosystem that emerged around embedded phishing training~\cite{rapid7,cofense,wombattraining,knowbe4} claims improvements due to the benefits of training in a recent collaborative report~\cite{verizon2019dbir}, but does not report results of experiment in controlled settings~\cite{siadati2017measuring}.

Phishing warnings have been studied extensively in the context of browsers (e.g.,~\cite{egelman2008you,akhawe2013alice}). Some recent works~\cite{petelka2019put} also evaluate different kinds of warnings shown on the email client. While too frequent warnings can susceptible to habituation and lose some of their effectiveness over time~\cite{vance2017we}, the literature agrees that carefully-timed warnings are in general effective.

\paragraph{Crowd-sourced phishing detection}
Several companies already provide tools to report phishing emails, to quickly detect new attacks using aggregate information across multiple customers~\cite{rapid7,wombatbutton}. 
The same companies report that users are improving at reporting phishing attempts over time~\cite{verizon2019dbir,verizon2020dbir}, however, other work showed that users are reticent to report phishing to the IT because of the lack of transparency in the process~\cite{kwak2020users} and lack of fast responses from the system~\cite{williams2018exploring}.
Prior to our work, it was not known if employees as a crowd-sourcing mechanism in a closed scenario, such as a corporation that manages reported phishing in-house, works effectively with acceptable operational workload.
Few recent works also suggest this concept, but do not evaluate it~\cite{burda2020don, nguyen2021comparison}.

%% file: sections/conclusion.tex
Thanks to our long-term and large-scale experiment, in this paper we supported several prior findings such as effectiveness of warnings with increased ecological validity. Further, we found that embedded phishing training, as commonly deployed in the industry today, is not effective and can in fact have negative side effects. In this regard, our results contradict prior literature and common industry practices. Finally, we are the first to experimentally demonstrate that crowd-sourced phishing detection is effective and practical in a single organization. 

Based on these results, we encourage organizations to adopt phishing prevention tools like warnings that have been extensively studied and where the available literature supports their effectiveness overwhelmingly. We call for caution in the deployment of methods like embedded phishing exercises and training, where the the existing literature is less unanimous about their effectiveness, and our research discovers potential negative side effects. We recommend organizations to consider crowd-sourced phishing detection as a new and complementary way to improve the overall phishing prevention capabilities of the organization, since its effectiveness looks promising and operational workload remains low.

Our work also identifies topics where more research is needed. Our research shows that the effectiveness of phishing exercise and training has not been sufficiently measured, and it remains unknown what is the most effective way to deliver embedded phishing training. More research is also needed to better understand the (psychological) effects of phishing exercises and training that is embedded into the normal working context of employees, and how such effects may influence the employees' future handling of real phishing emails.

%% file: sections/additional_results.tex
We report here some of the materials we used in our experiment: the simulated phishing emails, the embedded training webpage that participants viewed when performing the dangerous action on the emails, and the questionnaire we administered at the end of the study.

Due to space limitations, we only report a sample of each material: one embedded training webpage tailored for a specific simulated email; four simulated phishing emails, and the questionnaire questions that gave us some insights that we reported in this paper.
We believe this sample is sufficient to understand our design in full details.

\subsection{Embedded Training Webpage}
\label{app:training}

\begin{figure*}[b]
	\centering
	\includegraphics[width=0.8\linewidth]{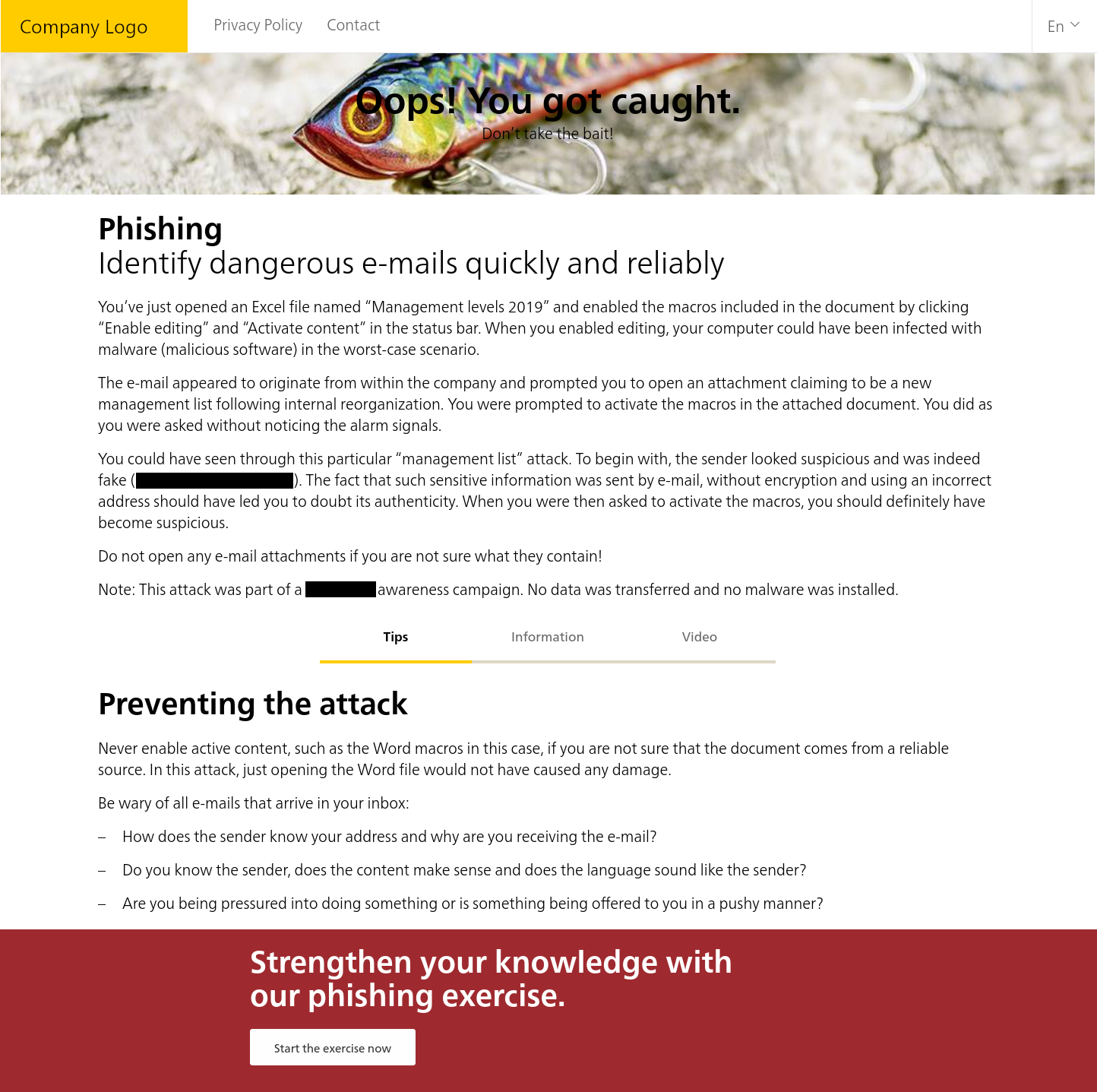}
	\caption{Sample contextual training webpage displayed to employees after dangerous actions on a simulated phishing email.}
	\label{fig:training_page}
\end{figure*}

We show in Figure~\ref{fig:training_page} the contextual training webpage displayed when employees performed the dangerous action of one simulated phishing email. 
It contains tailored information, explanation about the awareness campaign, and the tabs further contain information about email threats and an instructional video.

\subsection{Simulated Phishing Emails}
\label{app:emails}

\begin{figure*}[!t]
	\centering
	\begin{subfigure}[]{0.45\linewidth}
		\centering
		\includegraphics[width=\linewidth]{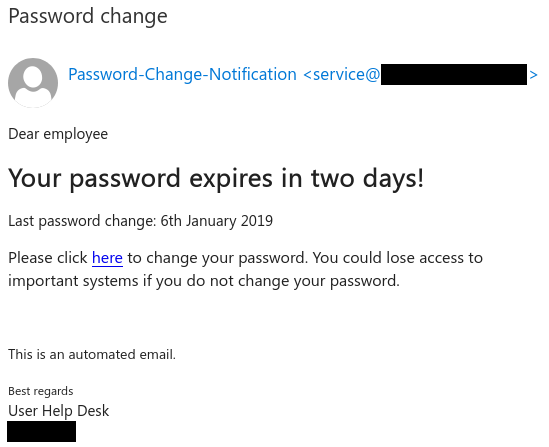}
		\caption{Email prompting to change the organization's password.}
		\label{fig:email1}
	\end{subfigure}
	~
	\begin{subfigure}[]{0.45\linewidth}
	\centering
	\includegraphics[width=\linewidth]{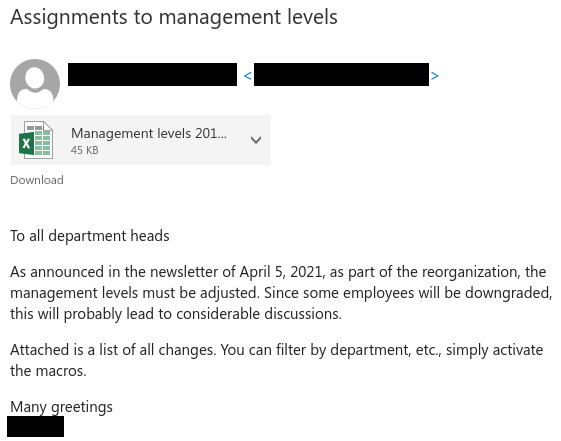}
	\caption{Email with attachment with malicious macros.}
	\label{fig:email2}
	\end{subfigure}
	\\[5em]
	\begin{subfigure}[b]{0.45\linewidth}
	\centering
	\includegraphics[width=\linewidth]{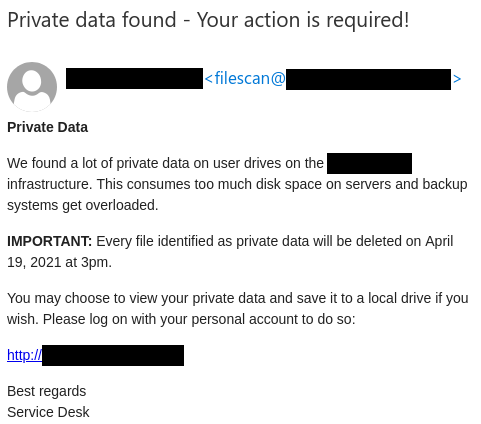}
	\caption{Email prompting to check one's files in a corporate drive.}
	\label{fig:email3}
	\end{subfigure}
	~
	\begin{subfigure}[b]{0.45\linewidth}
	\centering
	\includegraphics[width=\linewidth]{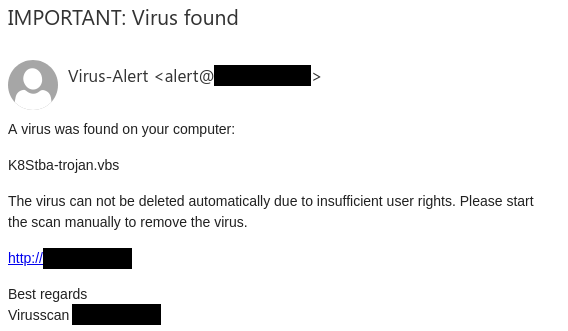}
	\caption{Email alerting of presumed malware.}
	\label{fig:email4}
	\end{subfigure}
	\caption{Sample of the simulated phishing emails. We report three emails containing a link to a malicious webpage (e.g., that asked for credentials, or prompted a download), and one with a malicious attachment.}
	\label{fig:emails}
\end{figure*}

Figure~\ref{fig:emails} shows four of the simulated phishing emails we sent to participants.
The reported emails either aim to get the participants to click a link to a malicious webpage, which in turn aims to get the participants to do the unsafe action (e.g., submit their credentials), or aim to get the participant to download an attachment, e.g., a document that prompts to enable macros.
Each email used different triggers to urge participants to click, such as curiosity or scare of consequences.

\subsection{Questionnaire}
\label{app:questionnaire}

The questionnaire consisted of 27 closed-ended questions, such as yes/no questions, and multiple choice questions where participants could select more than one answer.
Every question offered an answer such as not knowing or not remembering, and could also be left unanswered by the respondent.
We can group the questions in five main categories:
\begin{itemize}
	\item Familiarity with computer and email security threats.
	\item Email warnings.
	\item The button to report phishing in the email client.
	\item Remembering suspicious emails and security incidents.
	\item The contextual training webpage.
\end{itemize}
The groups of questions about email warnings, the report button, and the contextual training webpage were preceded by a recall about the tool, e.g., we displayed a screenshot of the training webpage immediately before its questions.

The questions about the deployed tools (email warnings, report button, contextual training webpage) were preceded by recalling it to respondents, e.g., we displayed a screenshot of the training webpage immediately before asking questions about it.
They asked the respondent if they remembered noticing the tool in the past 12 months and using it, and what they thought about the tool.

We report as a sample the questions about the training page in the following.

\begin{itemize}
	\item \texttt{Q22}: Do you remember seeing this training page during the past 12 months?
	\begin{itemize}
		\footnotesize
		\item \textit{Yes; No; I'm not sure}
	\end{itemize}
	\item \texttt{Q23}: How did you feel when you saw the training page?
	\begin{itemize}
		\footnotesize
		\item \textit{Embarrassed. I understood that I had made a mistake.}
		\item \textit{Concerned. I realized I had endangered my own and my company’s online security.}
		\item \textit{Safe. I felt that the organization is protecting me online.}
		\item \textit{Uninterested. Seeing the training page did not trigger any emotional reaction.}
		\item \textit{I don’t remember.		}
	\end{itemize}
	\item \texttt{Q24}: How much time would you estimate you spent on the training page?
	\begin{itemize}
		\footnotesize
		\item \textit{More than a minute. I read the whole page carefully.}
		\item \textit{Less than a minute. I briefly skimmed the provided information.}
		\item \textit{Few seconds. I opened the page but did not read its contents.}
		\item \textit{I don’t remember.}
	\end{itemize}
	\item \texttt{Q25}: Did you find the content of the training page trustworthy?
	\begin{itemize}
		\footnotesize
		\item \textit{Yes. I thought it was from a legitimate source like the IT department of the organization.}
		\item \textit{No. The training page looked suspicious to me (perhaps a scam).}
		\item \textit{I don’t remember.		}
	\end{itemize}
\item \texttt{Q26}: Did you find the content of the training page useful?
\begin{itemize}
	\footnotesize
	\item \textit{Yes. I found it a good reminder of threats of malicious emails. }
	\item \textit{No. The provided information was not helpful to me.}
	\item \textit{Not sure.}
\end{itemize}
\item \texttt{Q27}: After visiting the training page, did your attitude towards suspicious emails change?
\begin{itemize}
	\footnotesize
	\item \textit{Yes. I learned more about how to check suspicious emails.}
	\item \textit{Yes. I realized that suspicious emails can be part of a corporate training campaign. }
	\item \textit{Yes. I felt that the organization is protecting me from bad emails.}
	\item \textit{No. I already knew the information in the webpage.}
	\item \textit{No. The content of the page was not clear or informative.}
	\item \textit{Don’t remember.}
\end{itemize}
\end{itemize}